\documentclass{article}

\usepackage[pdftex,breaklinks,colorlinks,
citecolor=blue,
urlcolor=blue]{hyperref}
\usepackage{float}
\usepackage{multicol}
\usepackage{subfigure}
\usepackage{tabularx}
\usepackage{booktabs}
\usepackage{graphicx}
\usepackage{amsmath}
\usepackage{amssymb}
\usepackage{latexsym}
\usepackage{mathrsfs}
\usepackage{geometry}
\usepackage{chngcntr}
\counterwithin{figure}{section}
\counterwithin{table}{section}
\numberwithin{equation}{section}

\usepackage{authblk}
\title{Scaling analyses of the spectral dimension in $3$-dimensional causal dynamical triangulations}
\author[1]{Joshua H. Cooperman}
\affil[1]{15 Pond Drive West, Rhinebeck, New York 12572, United States}
\begin{document}

\maketitle

\begin{abstract}
The spectral dimension measures the dimensionality of a space as witnessed by a diffusing random walker. Within the causal dynamical triangulations approach to the quantization of gravity \cite{JA&JJ&RL1,JA&JJ&RL2,JA&RL}, the spectral dimension exhibits novel scale-dependent dynamics: reducing towards a value near $2$ on sufficiently small scales, matching closely the topological dimension %of $4$ 
on intermediate scales, and decaying %exponentially
in the presence of positive curvature on sufficiently large scales \cite{JA&JJ&RL7,JA&JJ&RL6,DB&JH,JHC2,JHC&KL&JMM,DNC&JJ,RK}. %Although the first measurement of spectral dimension within causal dynamical triangulations dates back more than a decade, a comprehensive scaling analysis of the spectral dimension is still lacking. 
I report the first comprehensive scaling analysis of the small-to-intermediate scale spectral dimension for the test case of the causal dynamical triangulations of $3$-dimensional Einstein gravity. 
%I report on a finite-size scaling analysis of the spectral dimension %for relatively short diffusion times 
%within the causal dynamical triangulations of $(2+1)$-dimensional Einstein gravity. % for spherical spatial topology. 
I find that the spectral dimension scales trivially with the diffusion constant. %, that the diffusion time scales anomalously with the number of $3$-simplices on sufficiently small scales, 
I find that the spectral dimension is completely finite in the infinite volume limit, and I argue that its maximal value %approaches a finite value 
is exactly consistent with the topological dimension of $3$ in this limit. I find that the spectral dimension reduces further towards a value near $2$ as this case's bare coupling approaches its phase transition, and I present evidence against the conjecture that the bare coupling simply sets the overall scale of the quantum geometry \cite{JA&JJ&RL3}. On the basis of these findings, I advance a tentative physical explanation for the dynamical reduction of the spectral dimension observed within causal dynamical triangulations: branched polymeric quantum geometry on sufficiently small scales. My analyses should facilitate attempts to employ the spectral dimension as a physical observable with which to delineate renormalization group trajectories in the hope of taking a continuum limit of causal dynamical triangulations at a nontrivial ultraviolet fixed point \cite{JA&DNC&JGS&JJ,JA&AG&JJ&AK&RL,JHC,JHC4,DNC&JJ}. 
\end{abstract}

\section{Introduction}

%The $4$-dimensionality of the our universe is an observational fact. In most of physics, this fact is taken for granted and input as a fixed parameter. In some approaches to quantum gravity, however, this is not so: one hopes to find that the theory predicts $4$-dimensionality on the scales that we have thus far observed. As Carlip in particular has discussed, How does one measure dimensionality? There are various ways. 

%The spacetime dimension of a superposition of geometries is not necessarily the spacetime dimension of those geometries. Moreover, the dimension as measured by probing a geometry at one length scale is not necessarily the same as the dimension as measured by probing the geometry at another scale. 

The causal dynamical triangulations approach to the quantization of gravity produced the first significant indication that spacetime dimensionality dynamically reduces on sufficiently small length scales \cite{JA&JJ&RL7}. Specifically, numerical measurements revealed the spectral dimension decreasing from a value of approximately $4$ on intermediate scales to a value of approximately $2$ on sufficiently small scales \cite{JA&JJ&RL7,JA&JJ&RL6,RK}. %of order $10\ell_{\mathrm{P}}$ 
More recent such measurements, performed at different values of the bare couplings, showed dynamical reduction towards a value closer to $\frac{3}{2}$ \cite{DNC&JJ}. %---on scales estimated to be close to the Planck scale. 
Evidence for dynamical dimensional reduction has subsequently emerged from several other approaches to the quantization of gravity \cite{SC}. 

Although the original paper documenting dynamical dimensional reduction % in causal dynamical triangulations 
dates back more than a decade \cite{JA&JJ&RL7}, only recently did Coumbe and Jurkiewicz perform a more thorough study \cite{DNC&JJ}. %there is still no comprehensive scaling analysis of the spectral dimension in the literature on causal dynamical triangulations. 
I now report a comprehensive study, including several scaling analyses not considered by Coumbe and Jurkiewicz, for the case of the causal dynamical triangulations of $3$-dimensional Einstein gravity. This computationally simpler model reproduces all of the known phenomenology of the causal dynamical triangulations of $4$-dimensional Einstein gravity, excepting certain aspects of the latter model's phase structure \cite{JA&JJ&RL3,CA&SC&JHC&PH&RKK&PZ,DB&JH,DB&JH2,JHC2,JHC&KL&JMM,JHC&JMM,RK}. Indeed, as the results reported below demonstrate, the former model's spectral dimension exhibits the behavior of the latter model's spectral dimension even more closely than previously appreciated. I study the $3$-dimensional model as an informative prelude to the more realistic $4$-dimensional model. 

In the case of the causal dynamical triangulations of $3$-dimensional Einstein gravity, numerical measurements of the spectral dimension involve three parameters: the diffusion constant $\rho$ characterizing random walks on causal triangulations, the number $N_{3}$ of $3$-simplices used to construct causal triangulations, and the bare coupling $k_{0}$ of the Euclidean Regge action for causal triangulations. After introducing the causal dynamical triangulations approach in section \ref{CDT} and the spectral dimension in section \ref{specdim}, I study the dependence of the spectral dimension on each of these parameters in section \ref{FSS}. Investigating variation of the diffusion constant $\rho$ in subsection \ref{diffconstdependence}, I show that the diffusion constant $\rho$ trivially rescales the spectral dimension's diffusion time dependence. Investigating variation of the number $N_{3}$ of $3$-simplices in subsection \ref{N3dependence}, I show that the spectral dimension remains finite in the infinite volume limit, as previously expected \cite{JA&JJ&RL7,JA&JJ&RL6,DNC&JJ}, but slightly overshoots the topological dimension of $3$, as previously observed \cite{DB&JH}. Investigating variation of the bare coupling $k_{0}$ in subsection \ref{k0dependence}, I show that the amount of dimensional reduction increases towards the model's phase transition, just as occurs within the $4$-dimensional model \cite{DNC&JJ}. I also show in subsection \ref{k0dependence} that numerical measurements of the spectral dimension for different values of $k_{0}$ do not differ simply by rescaling as a function of $k_{0}$, casting doubt on the conjecture of Ambj\o rn, Jurkiewicz, and Loll that $k_{0}$ simply sets an overall scale \cite{JA&JJ&RL3}. I discuss consequences of the analyses of section \ref{FSS} in section \ref{conclusion}. I propose a way in which to reconcile the spectral dimension with the topological dimension in the infinite volume limit. I hypothesize a physical explanation for the spectral dimension's dynamical reduction. I argue that, beyond further characterizing the spectral dimension and its dynamical reduction, the analyses of section \ref{FSS} should prove valuable for employing the spectral dimension to delineate renormalization group flows \cite{JA&DNC&JGS&JJ,JHC,JHC4,DNC&JJ}. Following the treatment of \cite{JHC2}, I provide an explicit explanation of my method for estimating the spectral dimension and its error in appendix \ref{appendix}.

\section{Causal dynamical triangulations}\label{CDT}

Causal dynamical triangulations is an approach to the nonperturbative quantization of classical theories of gravity based on a particular lattice regularization of the path integral \cite{JA&JJ&RL1,JA&JJ&RL2,JA&RL}. Within this approach one studies the regularized gravitational transition amplitudes
\begin{equation}\label{causalpathsum}
\mathcal{A}_{\Sigma}[\Gamma]=\sum_{\substack{\mathcal{T}_{c}\cong[0,1]\times\Sigma \\ \mathcal{T}_{c}|_{\partial\mathcal{T}_{c}}=\Gamma}}\mu(\mathcal{T}_{c})\,e^{i\mathcal{S}_{\mathrm{cl}}[\mathcal{T}_{c}]/\hbar}
\end{equation} 
defined \emph{via} path summation over causal triangulations $\mathcal{T}_{c}$. 
A causal triangulation is a piecewise-Minkowski simplicial manifold possessing a global foliation by spacelike hypersurfaces all of the topology $\Sigma$; accordingly, a causal triangulation has the spacetime topology $[0,1]\times\Sigma$. One constructs a causal triangulation by appropriately gluing together $N_{D}$ causal $D$-simplices, each a simplicial piece of $D$-dimensional Minkowski spacetime with spacelike edges of squared invariant length $a^{2}$ and timelike edges of squared invariant length $-\alpha a^{2}$. $a$ is the lattice spacing, and $\alpha$ is a positive constant. %I depict the three types of causal $3$-simplices in figure \ref{3simplices}.
%\begin{figure}[h]
%\includegraphics[scale=0.5]{all_simplices_version_3.png}
%\caption{The three types of $3$-simplices employed in $3$-dimensional causal dynamical triangulations: from left to right, a $(3,1)$ $3$-simplex, a $(2,2)$ $3$-simplex, and a $(1,3)$ $3$-simplex}
%\label{3simplices}
%\end{figure}
Each causal $D$-simplex spans two adjacent spacelike hypersurfaces of the global foliation. %, represented in figure \ref{3simplices} by grey surfaces. 
One specifies $\mathcal{A}_{\Sigma}[\Gamma]$ %of equation \eqref{causalpathsum} 
by choosing the triangulation $\Gamma$ of the boundary $\partial\mathcal{T}_{c}$ of $\mathcal{T}_{c}$, setting the measure $\mu(\mathcal{T}_{c})$ equal to the inverse of the order of the automorphism group of $\mathcal{T}_{c}$, and forming the translation $\mathcal{S}_{\mathrm{cl}}[\mathcal{T}_{c}]$ of the classical action $S_{\mathrm{cl}}[\mathbf{g}]$ into the Regge calculus of causal triangulations. 

In the case of three dimensions, analytic calculations of $\mathcal{A}_{\Sigma}[\Gamma]$ are not currently possible. To study the quantum theory of gravity defined by the transition amplitudes of equation \eqref{causalpathsum}, 
one therefore employs numerical techniques, specifically Monte Carlo methods. To enable the application of such methods, one performs a Wick rotation of each causal triangulation by analytically continuing $\alpha$ to $-\alpha$ through the lower-half complex plane. One then numerically generates an ensemble of causal triangulations representative of those contributing to the partition function
\begin{equation}\label{partitionfunction}
\mathcal{Z}_{\Sigma}[\Gamma]=\sum_{\substack{\mathcal{T}_{c}\cong[0,1]\times\Sigma \\ \mathcal{T}_{c}|_{\partial\mathcal{T}_{c}}=\Gamma}}\mu(\mathcal{T}_{c})\,e^{-\mathcal{S}_{\mathrm{cl}}^{(\mathrm{E})}[\mathcal{T}_{c}]/\hbar},
\end{equation} 
obtained \emph{via} the Wick rotation from the path summation in equation \eqref{causalpathsum}. $\mathcal{S}_{\mathrm{cl}}^{(\mathrm{E})}[\mathcal{T}_{c}]$ is the resulting Euclidean action. Typically, one runs numerical simulations at a fixed number $N_{D}$ of causal $D$-simplices and a fixed number $T$ of leaves of the global foliation. %One estimates the expectation value of a physical observable $\mathcal{O}$ in the quantum state \eqref{causalpathsum} as the average $\langle\mathcal{O}\rangle$ over such an ensemble. 
Taking $S_{\mathrm{cl}}[\mathbf{g}]$ as that of $3$-dimensional Einstein gravity for $2$-sphere spatial topology $\Sigma$ and transforming $[0,1]$ into the $1$-sphere by periodic identification, %the Wick rotation yields the Euclidean action
\begin{equation}\label{CDTReggeaction}
\mathcal{S}_{\mathrm{cl}}^{(E)}[\mathcal{T}_{c}]=-k_{0}N_{0}+k_{3}N_{3}
\end{equation}
in which the bare couplings $k_{0}$ and $k_{3}$ are functions of the bare Newton constant $G_{0}$, the bare cosmological constant $\Lambda_{0}$, and the lattice spacing $a$, and $N_{0}$ is the number of $0$-simplices \cite{JA&JJ&RL2}. The numerical value of $\alpha$ is irrelevant in three dimensions; moreover, $k_{3}$ is not independent of $k_{0}$ as the partition function \eqref{partitionfunction} for fixed $N_{3}$ is only well-defined at the critical value $k_{3}^{c}(k_{0},N_{3})$. 

\begin{figure}[h]
\centering
\includegraphics[scale=0.55]{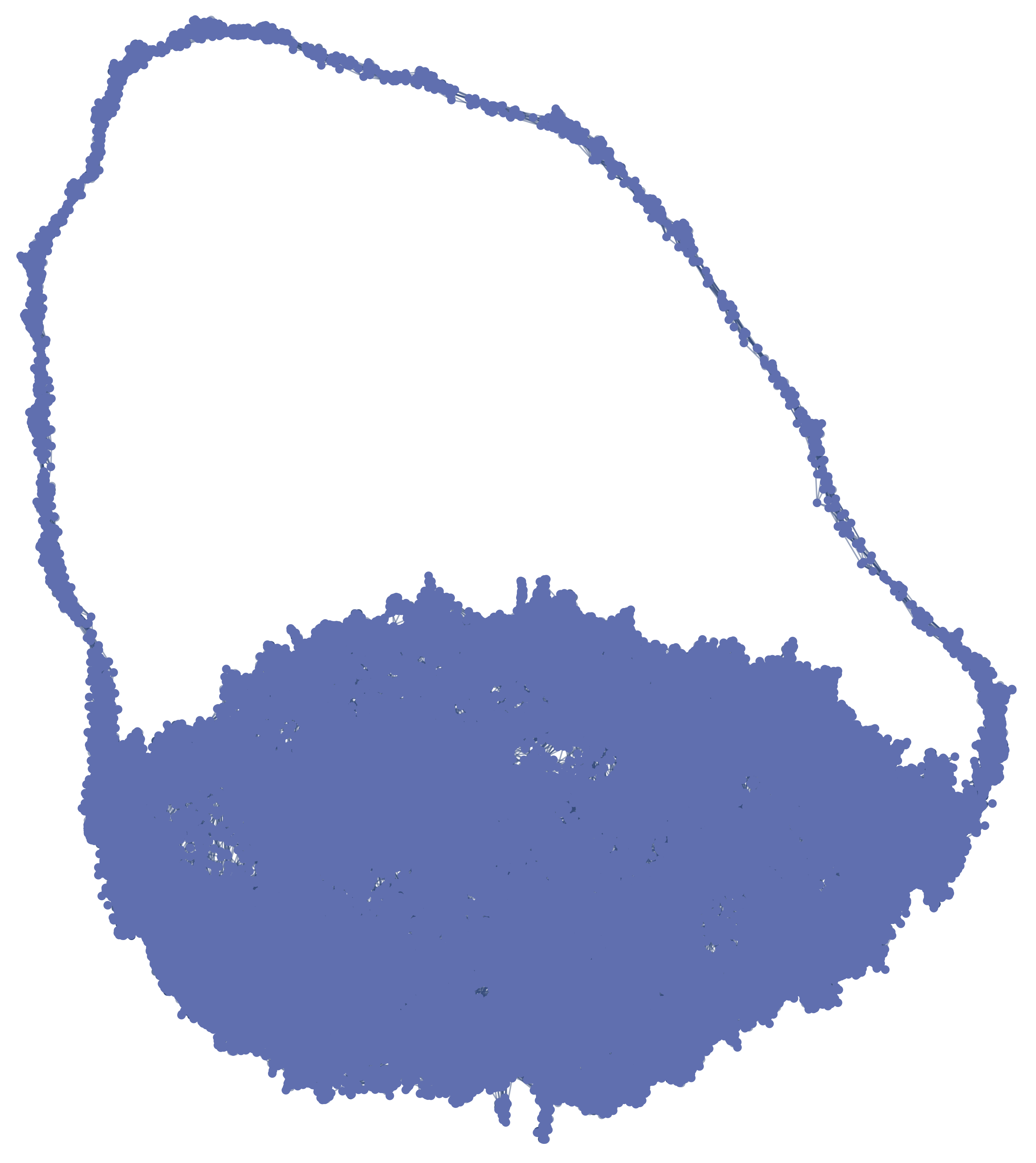}
\caption{A graphical representation of a dual causal triangulation within phase C. The global foliation advances around the loop; the condensate and the stalk are self-evident.}
\label{dualCT}
\end{figure}

For the action \eqref{CDTReggeaction} the partition function \eqref{partitionfunction} exhibits two phases of quantum geometry separated by a first-order transition at the critical value $k_{0}^{c}\approx3.3$ of $k_{0}$ \cite{JA&JJ&RL3,RK}.\footnote{As implemented numerically, my definition of $k_{0}$ differs by a factor of $2$ from that of \cite{JA&JJ&RL3}.} I consider exclusively the so-labeled phase C, a condensate of $3$-simplices that exhibits de Sitter-like properties on sufficiently large scales and dynamical reduction of the spectral dimension on sufficiently small scales \cite{JA&JJ&RL3,CA&SC&JHC&PH&RKK&PZ,DB&JH,DB&JH2,JHC2,JHC&KL&JMM,JHC&JMM,RK}. I consider only the condensate itself, excising the so-called stalk---a component of each causal triangulation now recognized as a numerical artifact \cite{JHC&JMM}---prior to numerical measurement of the spectral dimension. I provide a graphical representation of the condensate and the stalk in figure \ref{dualCT}. 
To distinguish the condensate from the stalk, I analyze the evolution of the discrete spatial $2$-volume in the global foliation as described, for instance, in \cite{CA&SC&JHC&PH&RKK&PZ,JHC&JMM}. I return to the other phase, so-labeled A, in section \ref{conclusion}.

Ultimately, one hopes to remove the lattice regularization of causal triangulations by finding a continuum limit %of $\mathcal{A}_{\Sigma}[\Gamma]$ 
in which the lattice spacing vanishes while expectation values of physical observables remain finite. Typically, the existence of such a continuum limit is contingent upon the existence of an ultraviolet fixed point of the renormalization group, and the existence of such a fixed point is contingent upon the existence of a second-order (or higher-order) phase transition. Since the causal dynamical triangulations of $3$-dimensional Einstein gravity lacks such phase transitions, one would not expect this model to possess a continuum limit. Ambj\o rn, Jurkiewicz, and Loll have argued, nevertheless, that this model does possess a continuum limit and, moreover, that taking this continuum limit requires no tuning of $k_{0}$ \cite{JA&JJ&RL3}. I return to their argument in subsection \ref{k0dependence}. I explain in section \ref{conclusion} that the analyses of the spectral dimension reported in section \ref{FSS} will in any case prove instructive for attempts to find a continuum limit of the causal dynamical triangulations of $4$-dimensional Einstein gravity, a model that possess at least one second-order phase transition \cite{JA&DNC&JGS&AG&JJ,JA&DNC&JGS&AG&JJ&NK&RL,JA&SJ&JJ&RL1,JA&SJ&JJ&RL2}. Furthermore, one hopes to demonstrate an Osterwalder-Schrader-type theorem for causal dynamical triangulations to understand the import of numerical results derived from the partition function \eqref{partitionfunction} for the transition amplitude \eqref{causalpathsum}.

\section{Spectral dimension}\label{specdim}

Several measures of dimensionality, including the so-called spectral dimension, are based on properties of random walks. Consider a random walker diffusing through a $D$-dimensional Riemannian manifold $\mathcal{M}$ with metric $\mathbf{g}$. The heat equation,
\begin{equation}\label{heatequation}
\frac{\partial}{\partial s}K(\mathbf{x},\mathbf{x}',s)=\wp\nabla_{\mathbf{g}}^{2}K(\mathbf{x},\mathbf{x}',s),
\end{equation}
governs the random walker's diffusion. Subject to the initial condition
\begin{equation}
K(\mathbf{x},\mathbf{x}',0)=\sqrt{\mathrm{det}\,\mathbf{g}(\mathbf{x})}\,\delta^{(D)}(\mathbf{x}-\mathbf{x}')
\end{equation} 
specifying the walker's initial location, $K(\mathbf{x},\mathbf{x}',s)$ is the probability of diffusion from location $\mathbf{x}$ to location $\mathbf{x}'$ in diffusion time $s$ for diffusion constant $\wp$. The return probability, 
\begin{equation}
P(s)=\frac{1}{\mathrm{vol}(\mathcal{M})}\int_{\mathcal{M}}\mathrm{d}^{D}\mathbf{x}\sqrt{\mathrm{det}\,\mathbf{g}(\mathbf{x})}\,K(\mathbf{x},\mathbf{x},s),
\end{equation}
is the probability that the walker returns to its initial location in diffusion time $s$ (irregardless of its initial location). One defines the spectral dimension $D_{\mathfrak{s}}(s)$ as follows:
\begin{equation}
D_{\mathfrak{s}}(s)=-2\frac{\mathrm{d}\ln{P(s)}}{\mathrm{d}\ln{s}}.
\end{equation}
$D_{\mathfrak{s}}(s)$ quantifies the scaling of $P(s)$ with $s$; the factor of $-2$ ensures that 
\begin{equation}\label{specdimlimit0}
\lim_{s\rightarrow0}D_{\mathfrak{s}}(s)=D,
\end{equation}
yielding agreement of $D_{\mathfrak{s}}(0)$ with the topological dimension $D$. The content of equation \eqref{specdimlimit0} constitutes a primary reason for regarding $D_{\mathfrak{s}}(s)$ as a measure of dimension. %The content of equation \eqref{specdimlimit0} is also expected: 
Of course, any scale-dependent measure of dimension should yield $D$ for $s=0$ since a sufficiently small neighborhood of a Riemannian manifold is isomorphic to Euclidean space. A sufficiently large neighborhood of a Riemannian manifold %, provided that $\mathbf{g}$ possesses nonzero curvature, 
is generically not isomorphic to Euclidean space, in which case $D_{\mathfrak{s}}(s)\neq D$ for $s\neq 0$. $D_{\mathfrak{s}}(s)$ provides a measure of the dimensionality of $\mathcal{M}$ on the scale specified by $s$.\footnote{In the mathematics literature the spectral dimension is typically defined as $\lim_{s\rightarrow0}D_{\mathfrak{s}}(s)$.}

I now adapt the above definition of the spectral dimension to the setting of causal triangulations \cite{JA&JJ&RL7,JA&JJ&RL6,DB&JH}. Consider a random walker diffusing through a $D$-dimensional Wick-rotated causal triangulation $\mathcal{T}_{c}$. The integrated heat equation takes the form
\begin{equation}\label{discreteheatequation}
\mathcal{K}(\mathsf{s},\mathsf{s}',\sigma+1)=(1-\rho)\mathcal{K}(\mathsf{s},\mathsf{s}',\sigma)+\frac{\rho}{N(\mathscr{N}(\mathsf{s}'))}\sum_{\mathsf{s}''\in\mathscr{N}(\mathsf{s}')}\mathcal{K}(\mathsf{s},\mathsf{s}'',\sigma).
\end{equation}
Subject to the initial condition
\begin{equation}
\mathcal{K}(\mathsf{s},\mathsf{s}',0)=\delta_{\mathsf{s}\mathsf{s}'}
\end{equation}
specifying the walker's initial $D$-simplex, $\mathcal{K}(\mathsf{s},\mathsf{s}',\sigma)$ is the probability of diffusion from $D$-simplex $\mathsf{s}$ to $D$-simplex $\mathsf{s}'$ in $\sigma$ diffusion time steps for diffusion constant $\rho$. $\mathscr{N}(\mathsf{s}')$ is the set of $N(\mathscr{N}(\mathsf{s}'))$ $D$-simplices neighboring the simplex $\mathsf{s}'$. The return probability,
\begin{equation}
\mathcal{P}(\sigma)=\frac{1}{N_{3}(\mathcal{T}_{c})}\sum_{\mathsf{s}\in\mathcal{T}_{c}}\mathcal{K}(\mathsf{s},\mathsf{s},\sigma),
\end{equation}
is the probability that the walker returns to its initial $D$-simplex in $\sigma$ diffusion time steps (irregardless of its initial $D$-simplex). One defines the spectral dimension $\mathcal{D}_{\mathfrak{s}}(\sigma)$ as follows:
\begin{equation}\label{specdimTc}
\mathcal{D}_{\mathfrak{s}}(\sigma)=-2\frac{\mathrm{d}\ln{\mathcal{P}(\sigma)}}{\mathrm{d}\ln{\sigma}}
\end{equation}
for appropriate finite differences. In particular, expanding the logarithmic derivative in equation \eqref{specdimTc}, %in the definition of $D_{s}(\sigma)$, %since 
%\begin{equation}
%\frac{\mathrm{d}\ln{\mathcal{P}(\sigma)}}{\mathrm{d}\ln{\sigma}}=\frac{\sigma}{\mathcal{P}(\sigma)}\frac{\mathrm{d}\mathcal{P}(\sigma)}{\mathrm{d}\sigma},
%\end{equation}
\begin{equation}
\mathcal{D}_{\mathfrak{s}}(\sigma)=-\frac{\sigma}{\mathcal{P}(\sigma)}\left[\mathcal{P}(\sigma+1)-\mathcal{P}(\sigma-1)\right]
\end{equation}
to lowest order in $\Delta\sigma=1$.

I finally adapt the above definition of the spectral dimension to an ensemble of causal triangulations. Consider an ensemble of $N(\mathcal{T}_{c})$ $D$-dimensional Wick-rotated causal triangulations $\mathcal{T}_{c}$. The expectation value of $\mathcal{D}_{\mathfrak{s}}(\sigma)$ in the quantum state \eqref{partitionfunction} is defined as follows:
\begin{equation}
\mathbb{E}[\mathcal{D}_{\mathfrak{s}}(\sigma)]=\frac{1}{\mathcal{Z}[\Gamma]}\sum_{\substack{\mathcal{T}_{c}\cong[0,1]\times\Sigma \\ \mathcal{T}_{c}|_{\partial\mathcal{T}_{c}}=\Gamma}}\mu(\mathcal{T}_{c})\,e^{-\mathcal{S}^{(\mathrm{E})}_{\mathrm{cl}}[\mathcal{T}_{c}]/\hbar}\,\mathcal{D}_{\mathfrak{s}}^{(\mathcal{T}_{c})}(\sigma).
\end{equation}
I estimate $\mathbb{E}[\mathcal{D}_{\mathfrak{s}}(\sigma)]$ by the ensemble average 
\begin{equation}
\langle\mathcal{D}_{\mathfrak{s}}(\sigma)\rangle=\frac{1}{N(\mathcal{T}_{c})}\sum_{j=1}^{N(\mathcal{T}_{c})}\mathcal{D}_{\mathfrak{s}}^{(\mathcal{T}_{c}^{(j)})}(\sigma).
\end{equation}
$\langle \mathcal{D}_{s}(\sigma)\rangle$ provides a measure of the dimensionality of the quantum geometry on the scale corresponding to $\sigma$ defined by an ensemble of causal triangulations. I report numerical measurements of $\langle\mathcal{D}_{\mathfrak{s}}(\sigma)\rangle$ for several ensembles of causal triangulations within phase C in section \ref{FSS}. %Since I want to %estimate $\mathbb{E}[\mathcal{D}_{\mathfrak{s}}(\sigma)]$ 
%measure $\langle\mathcal{D}_{\mathfrak{s}}(\sigma)\rangle$ for the condensate, I excise the stalk from each causal triangulation before computing the spectral dimension. 

\section{Scaling analyses}\label{FSS}

Three parameters---the diffusion constant $\rho$, the number $N_{3}$ of $3$-simplices, and the bare coupling $k_{0}$---enter into a numerical measurement of the ensemble average spectral dimension $\langle\mathcal{D}_{\mathfrak{s}}(\sigma)\rangle$.\footnote{One also specifies the number $T$ of leaves of the global foliation, but the number of leaves within the condensate is dynamically determined, so the dependence of $\langle\mathcal{D}_{\mathfrak{s}}(\sigma)\rangle$ on $T$ is quite weak.} I explore the dependence of $\langle \mathcal{D}_{\mathfrak{s}}(\sigma)\rangle$ on $\rho$ for fixed $N_{3}$ and $k_{0}$ in subsection \ref{diffconstdependence}, the dependence of $\langle \mathcal{D}_{\mathfrak{s}}(\sigma)\rangle$ on $N_{3}$ for fixed $\rho$ and $k_{0}$ in subsection \ref{N3dependence}, and the dependence of $\langle \mathcal{D}_{\mathfrak{s}}(\sigma)\rangle$ on $k_{0}$ for fixed $\rho$ and $N_{3}$ in subsection \ref{k0dependence}. In each of these explorations, I look for evidence of scaling of $\langle \mathcal{D}_{\mathfrak{s}}(\sigma)\rangle$ with the parameter being varied. 

The exploration of subsection \ref{N3dependence} is a standard finite-size scaling analysis.
%In the context of numerical simulation of physical systems, one often measures physical observables at different values of the simulation's control parameters, and one can only measure physical observables at finite values of the simulation's control parameters. 
%One performs a finite-size scaling analysis to extrapolate such measurements towards infinite values of %the control parameters. In the context of causal dynamical triangulations, the number $N_{3}$. % of $3$-simplices is the relevant control parameter. 
I present numerical measurements of $\langle \mathcal{D}_{\mathfrak{s}}(\sigma)\rangle$ %One measures physical observables at 
for increasing values of $N_{3}$ from which I attempt to extrapolate towards the infinite volume limit $N_{3}\rightarrow\infty$. 
%The purpose of a finite-size scaling analysis is to extrapolate numerical results towards the infinite volume limit in this case given by $N_{3}\rightarrow\infty$. 
As I discuss further in subsection \ref{N3dependence} below, one only expects to obtain a nontrivial continuum limit in conjunction with the infinite volume limit. Accordingly, finiteness of $\langle \mathcal{D}_{\mathfrak{s}}(\sigma)\rangle$ in the infinite volume limit, which I determine through this finite-size scaling analysis, may be necessary for finiteness of $\langle \mathcal{D}_{\mathfrak{s}}(\sigma)\rangle$ in a continuum limit. % observables well-defined continuum limit  in the infinite volume limit to be necessary for good behavior in the continuum limit. 

\subsection{Dependence on the diffusion constant $\rho$}\label{diffconstdependence}

\begin{figure}[h]
\centering
\includegraphics[width=\linewidth]{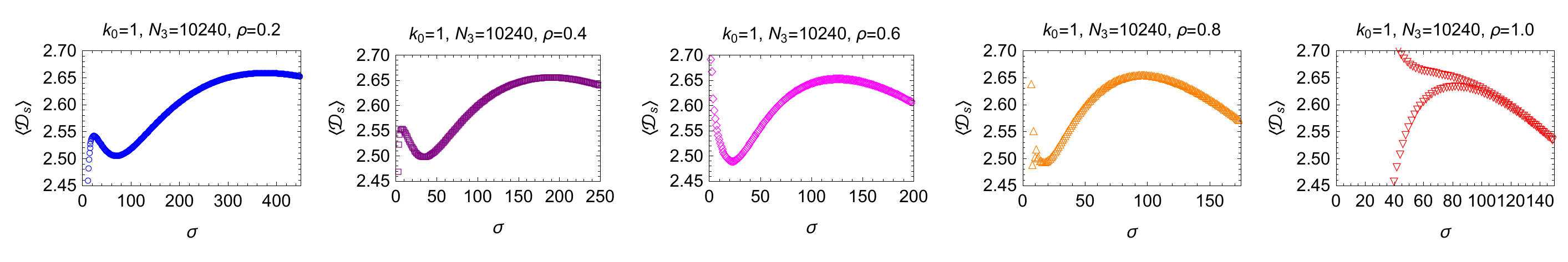}
%\subfigure[]{\includegraphics[width=0.2\linewidth]{specdimrho2.pdf}
%\label{specdimrho2}}
%\subfigure[]{\includegraphics[width=0.2\linewidth]{specdimrho4.pdf}
%\label{specdimrho4}}
%\subfigure[]{\includegraphics[width=0.2\linewidth]{specdimrho6.pdf}
%\label{specdimrho6}}
%\subfigure[]{\includegraphics[width=0.2\linewidth]{specdimrho8.pdf}
%\label{specdimrho8}}
%\subfigure[]{\includegraphics[width=0.2\linewidth]{specdimrho10.pdf}
%\label{specdimrho10}}
\caption{The ensemble average spectral dimension $\langle \mathcal{D}_{\mathfrak{s}}\rangle$ as a function of diffusion time $\sigma$ for diffusion constants $\rho=0.2$, $\rho=0.4$, $\rho=0.6$, $\rho=0.8$, and $\rho=1.0$ (from left to right) for $k_{0}=1.00$, $N_{3}=9267$, and $T=64$.}%\subref{} \rho=0.2 \subref{} \rho=0.4 \subref{} \rho=0.6 \subref{} \rho=0.8 \subref{} \rho=1.0}
\label{diffconstdependence10}
\end{figure}

I first explore the dependence of $\langle \mathcal{D}_{\mathfrak{s}}(\sigma)\rangle$ on $\rho$ holding fixed $k_{0}$ and $N_{3}$. Equation \eqref{heatequation} implies a trivial dependence: since the diffusion constant $\wp$---or, more precisely, $\wp^{-1}$---enters as a constant factor multiplying the diffusion time $s$, $\wp$ simply rescales $s$. This conclusion does not follow straightforwardly from equation \eqref{discreteheatequation}, and computing the expectation value of $\mathcal{D}_{\mathfrak{s}}(\sigma)$ with respect to the partition function \eqref{partitionfunction} further complicates predicting this dependence. %  leads one to expect a trivial dependence of $\langle \mathcal{D}_{s}(\sigma)\rangle$ on $\rho$: measurements of $\langle \mathcal{D}_{s}(\sigma)\rangle$ for different values of $\rho$ should be related by rescaling $\sigma$ by $\rho^{-1}$.  
I have instead performed measurements of $\langle\mathcal{D}_{\mathfrak{s}}(\sigma)\rangle$ for five values of $\rho$ at fixed $k_{0}$ and $N_{3}$ to determine its dependence on $\rho$. I display these measurements in figure \ref{diffconstdependence10}.

The plots in figure \ref{diffconstdependence10} exhibit the characteristic behavior of $\langle\mathcal{D}_{\mathfrak{s}}(\sigma)\rangle$ within phase C. The case of $\rho=0.8$ (orange, upright triangles) provides the clearest illustration. As $\sigma$ increases from a value of order $10$ to a value of order $100$, $\langle\mathcal{D}_{\mathfrak{s}}(\sigma)\rangle$ increases monotonically from a value of approximately $2.5$ to a value of approximately $2.65$. Read in reverse, this is the phenomenon of dynamical reduction of $\langle\mathcal{D}_{\mathfrak{s}}(\sigma)\rangle$, first reported in \cite{JA&JJ&RL7}. %(obscured in the case of $\rho=1.0$) 
While figure \ref{specdimN3} below attests that $\langle\mathcal{D}_{\mathfrak{s}}(\sigma)\rangle$ dynamically reduces to its minimal value $\langle\mathcal{D}_{\mathfrak{s}}\rangle_{\mathrm{min}}$ of approximately $2.5$ irregardless of the value of $N_{3}$, figures \ref{specdimvaryingk0sep}, \ref{specdimvaryingk0}, and \ref{specdimminvaryingk0} below attest that $\langle\mathcal{D}_{\mathfrak{s}}\rangle_{\mathrm{min}}$ depends on the value of $k_{0}$.\footnote{I exclude the (presumably) unphysically small values of $\langle\mathcal{D}_{\mathfrak{s}}(\sigma)\rangle$ for very small values of $\sigma$ ($\sigma<15$ for $\rho=0.2$, $\sigma<4$ for $\rho=0.4$, $\sigma<9$ for $\rho=0.8$).} %\footnote{$\langle\mathcal{D}_{\mathfrak{s}}(\sigma)\rangle$ dynamically reduces to a value of approximately $2$---or even $1.5$ depending on the relevant couplings---within phase C of the causal dynamical triangulations of $(3+1)$-dimensional Einstein gravity \cite{}.} 
Previous studies have attested \cite{JA&JJ&RL7,JA&JJ&RL6,DB&JH,JHC2,DNC&JJ} and figure \ref{specdimN3} below further attests that sufficiently small values of $N_{3}$ depress the maximal value $\langle\mathcal{D}_{\mathfrak{s}}\rangle_{\mathrm{max}}$ of $\langle\mathcal{D}_{\mathfrak{s}}(\sigma)\rangle$---in this case approximately $2.65$---below the value of the topological dimension---in this case $3$.\footnote{I exclude the (presumably) unphysically large values of $\langle\mathcal{D}_{\mathfrak{s}}(\sigma)\rangle$ for very small $\sigma$ ($\sigma<3$ for $\rho=0.6$, $\sigma<6$ for $\rho=0.8$, $\sigma<18$ for $\rho=1$).}  As $\sigma$ subsequently increases beyond a value of order $100$, $\langle\mathcal{D}_{\mathfrak{s}}(\sigma)\rangle$ decreases monotonically towards a value of $0$ \cite{DB&JH}. 
%For sufficiently small $\sigma$: for all values of $\rho$ except $1$, $\langle\mathcal{D}_{\mathfrak{s}}(\sigma)\rangle$ decreases from a value of approximately $2.65$ for $\sigma$ on the order of $10^{2}$ towards a value of approximately $2.5$ for $\sigma$ on the order of $10^{1}$ or $10^{0}$.  
In the cases of $\rho=0.2$, $\rho=0.4$, and $\rho=0.6$, $\langle\mathcal{D}_{\mathfrak{s}}(\sigma)\rangle$ exhibits additional structure for very small $\sigma$: as $\sigma$ increases from a value of order $1$ to a value of order $10$, $\langle\mathcal{D}_{\mathfrak{s}}(\sigma)\rangle$ decreases slightly from a value of approximately $2.55$ to a value of approximately $2.5$. I have yet to ascertain whether or not this decrease %over these values of $\sigma$ 
is a physical phenomenon; the physical explanation for the spectral dimension's dynamical reduction that I propose in section \ref{conclusion} implies that this decrease is not a physical phenomenon. 

\begin{figure}[h]
\centering
\includegraphics[width=0.45\linewidth]{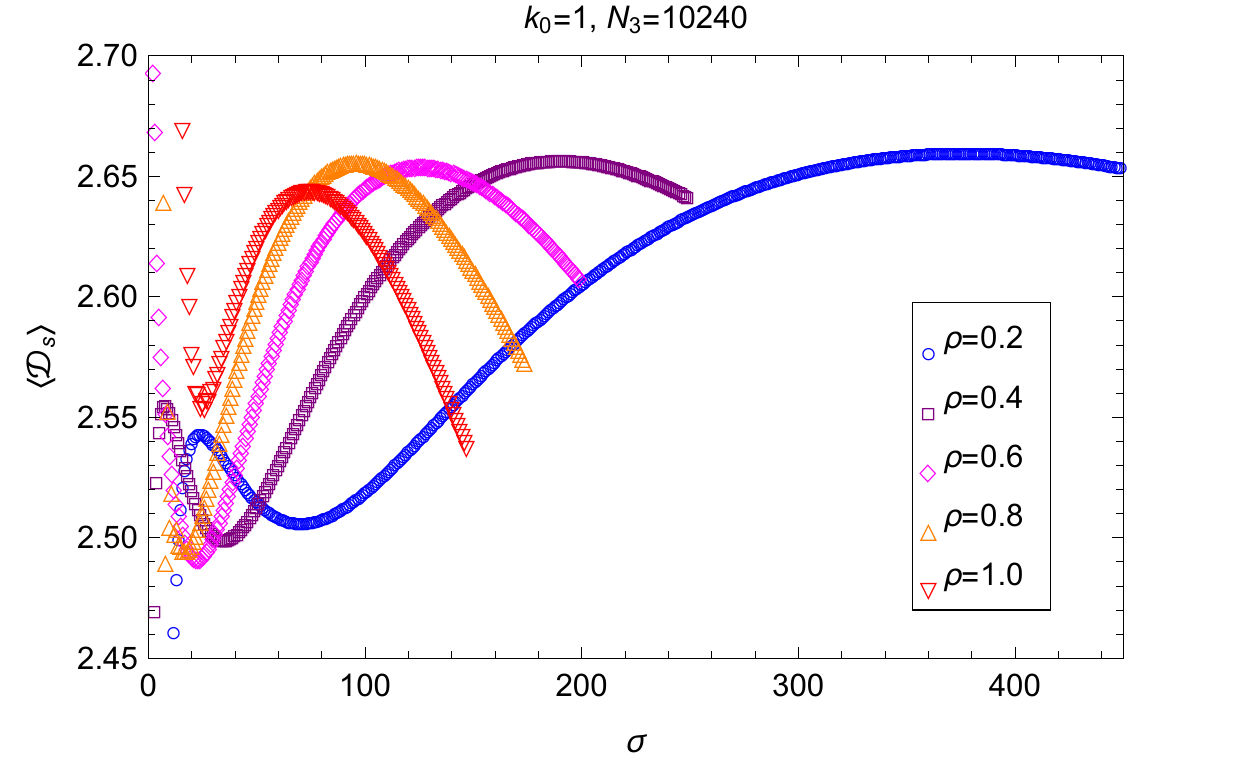}
\caption{The ensemble average spectral dimension $\langle \mathcal{D}_{\mathfrak{s}}\rangle$ as a function of diffusion time $\sigma$ for five values of the diffusion constant $\rho$ for $k_{0}=1.00$, $N_{3}=9267$, and $T=64$.}
\label{diffconstdependence10together}
\end{figure}

The dynamical reduction of $\langle\mathcal{D}_{\mathfrak{s}}(\sigma)\rangle$ is obscured for $\rho=1$ because random walks of even and odd $\sigma$ for $\sigma<100$ yield disparate estimates for $\langle\mathcal{D}_{\mathfrak{s}}(\sigma)\rangle$. As Benedetti and Henson first observed \cite{DB&JH}, and as figure \ref{diffconstdependence10} attests, choosing $\rho<1$ significantly smooths the oscillations in $\langle\mathcal{D}_{\mathfrak{s}}(\sigma)\rangle$. To render $\langle\mathcal{D}_{\mathfrak{s}}(\sigma)\rangle$ for $\rho=1$ useful for subsequent comparisons with $\langle\mathcal{D}_{\mathfrak{s}}(\sigma)\rangle$ for $\rho<1$, I compute an interpolated value $\langle\mathcal{D}_{\mathfrak{s}}(\sigma)\rangle_{\mathrm{int}}$ of $\langle\mathcal{D}_{\mathfrak{s}}(\sigma)\rangle$ for $\rho=1$ using a straightforward averaging technique:
\begin{equation}\label{interpolation}
\langle\mathcal{D}_{\mathfrak{s}}(\sigma)\rangle_{\mathrm{int}}=\frac{1}{4}\left[\langle\mathcal{D}_{\mathfrak{s}}(\sigma-1)\rangle+2\langle\mathcal{D}_{\mathfrak{s}}(\sigma)\rangle+\langle\mathcal{D}_{\mathfrak{s}}(\sigma+1)\rangle\right]\qquad (\rho=1).
\end{equation}
Alongside $\langle\mathcal{D}_{\mathfrak{s}}(\sigma)\rangle$ for $\rho=0.2
$, $\rho=0.4$, $\rho=0.6$, and $\rho=0.8$, I plot $\langle\mathcal{D}_{\mathfrak{s}}(\sigma)\rangle_{\mathrm{int}}$ (for $\rho=1$) in figure \ref{diffconstdependence10together}. 
The dynamical reduction of $\langle\mathcal{D}_{\mathfrak{s}}(\sigma)\rangle_{\mathrm{int}}$ is now evident.

\begin{figure}[h]
\centering
\includegraphics[width=0.45\linewidth]{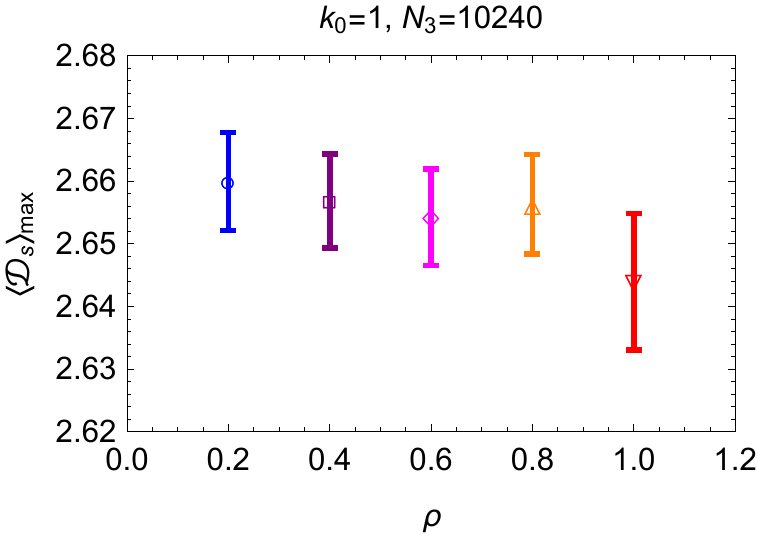}
\caption{The maximal value $\langle\mathcal{D}_{\mathfrak{s}}\rangle_{\mathrm{max}}$ of the ensemble average spectral dimension $\langle\mathcal{D}_{\mathfrak{s}}(\sigma)\rangle$ as a function of the diffusion constant $\rho$ for $k_{0}=1.00$, $N_{3}=9267$, and $T=64$.}
\label{specdimmaxvsrho}
\end{figure}

\begin{figure}[h]
\centering
\subfigure[]{
\includegraphics[width=0.45\linewidth]{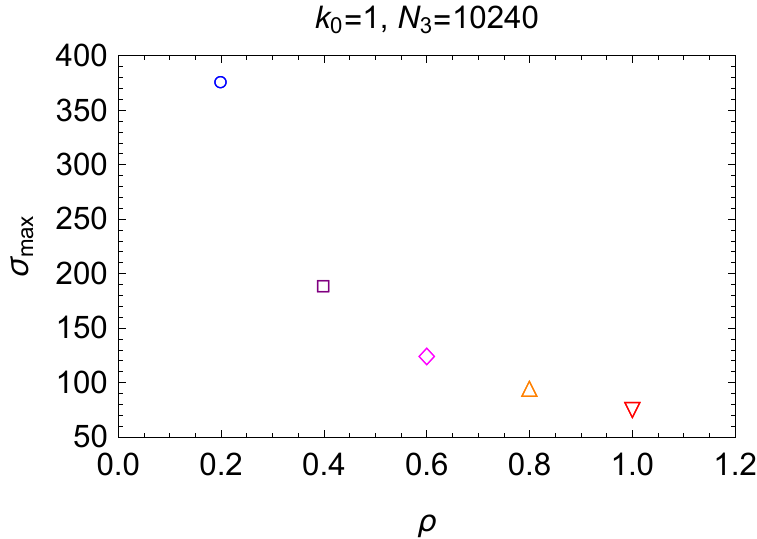}
\label{difftimespecdimmaxdiffconst10}}
\subfigure[]{
\includegraphics[width=0.45\linewidth]{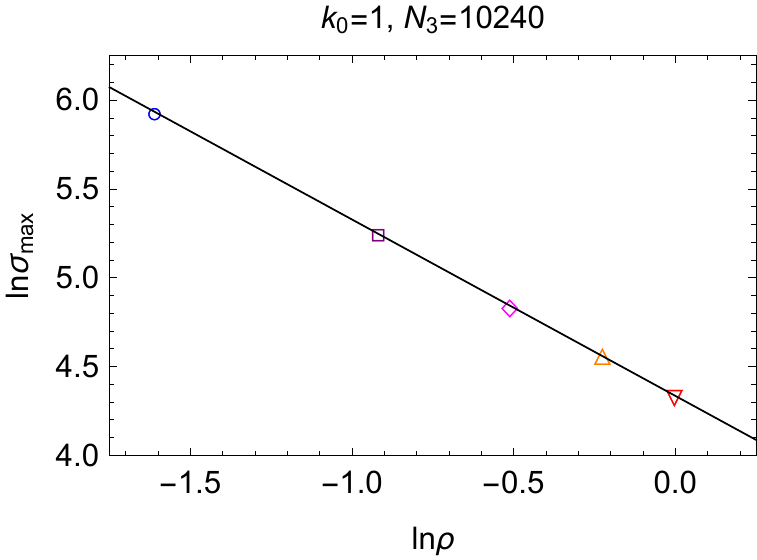}
\label{lndifftimespecdimmaxlndiffconst10}}
\label{difftimemaxdiffconst10}
\caption{\subref{difftimespecdimmaxdiffconst10} The diffusion time $\sigma_{\mathrm{max}}$ as a function of the diffusion constant $\rho$ for $k_{0}=1.00$, $N_{3}=9267$, and $T=64$. \subref{lndifftimespecdimmaxlndiffconst10} The logarithm $\ln{\sigma_{\mathrm{max}}}$ of the diffusion time $\sigma_{\mathrm{max}}$ as a function of the logarithm $\ln{\rho}$ of the diffusion constant $\rho$ for $k_{0}=1.00$, $N_{3}=9267$, and $T=64$ overlain with the linear least-squares fit $\ln{\sigma_{\mathrm{max}}}=4.33-0.99\ln{\rho}$.}
\end{figure}

\begin{figure}[h]
\centering
\includegraphics[width=0.45\linewidth]{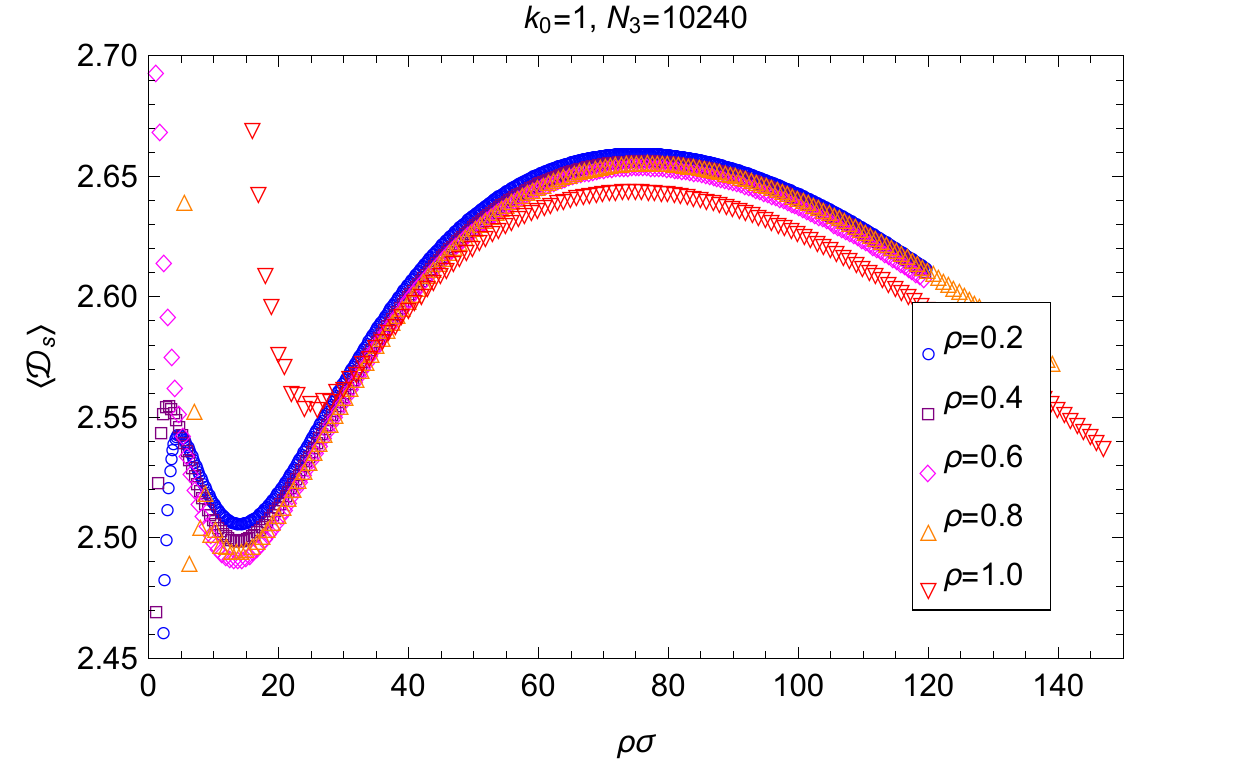}
\caption{The ensemble average spectral dimension $\langle \mathcal{D}_{s}\rangle$ as a function of the scaled diffusion time $\rho\sigma$ for five values of the diffusion constant $\rho$ for $k_{0}=1.00$, $N_{3}=9267$, and $T=64$.}
\label{specdimscaleddiffconst}
\end{figure}

I now turn to the dependence of $\langle\mathcal{D}_{\mathfrak{s}}(\sigma)\rangle$ on $\rho$. I wish to determine the relationship between measurements of $\langle\mathcal{D}_{\mathfrak{s}}(\sigma)\rangle$ performed for different values of $\rho$. I focus on %a clearly identifiable feature of $\langle\mathcal{D}_{\mathfrak{s}}(\sigma)\rangle$ for each value of $\rho$: 
the maximum of $\langle\mathcal{D}_{\mathfrak{s}}(\sigma)\rangle$ as a clearly identifiable point of comparison. I first plot $\langle\mathcal{D}_{\mathfrak{s}}\rangle_{\mathrm{max}}$ as a function of $\rho$ in figure \ref{specdimmaxvsrho},
which shows that $\langle\mathcal{D}_{\mathfrak{s}}\rangle_{\mathrm{max}}$ is a constant function of $\rho$ (up to measurement errors as computed in appendix \ref{appendix}). (Although this last statement includes $\langle\mathcal{D}_{\mathfrak{s}}\rangle_{\mathrm{max}}$ for $\rho=1$, I discount to some extent the value of $\langle\mathcal{D}_{\mathfrak{s}}\rangle_{\mathrm{max}}$ for $\rho=1$ since this value results from the naive interpolation \eqref{interpolation}.) %excluding the large values of $\langle\mathcal{D}_{\mathfrak{s}}(\sigma)\rangle$ for very small $\sigma$.
%To determine the dependence of $\langle\mathcal{D}_{\mathfrak{s}}(\sigma)\rangle$ on $\rho$, 
%As a distinguished point of comparison, I focus on $\sigma_{\mathrm{max}}$, the value of $\sigma$ at which $\langle \mathcal{D}_{\mathfrak{s}}(\sigma)\rangle$ attains its maximal value $\langle\mathcal{D}_{\mathfrak{s}}\rangle_{\mathrm{max}}$. 
I next plot $\sigma_{\mathrm{max}}$, %Denote by $\sigma_{\mathrm{max}}$ 
the value of $\sigma$ at which $\langle \mathcal{D}_{\mathfrak{s}}(\sigma)\rangle$ attains the value $\langle\mathcal{D}_{\mathfrak{s}}\rangle_{\mathrm{max}}$, as a function of $\rho$ in figure \ref{difftimemaxdiffconst10}\subref{difftimespecdimmaxdiffconst10}. %To determine the scaling of the $\sigma$-dependence of $\langle \mathcal{D}_{s}(\sigma)\rangle$ with $\rho$, I plot $\sigma_{\mathrm{max}}$ as a function of $\rho$ in figure 
In light of the regular decrease of $\sigma_{\mathrm{max}}$ with $\rho$ exhibited in figure \ref{difftimemaxdiffconst10}\subref{difftimespecdimmaxdiffconst10}, I test for scaling of $\sigma_{\mathrm{max}}$ with $\rho$ by plotting $\ln{\sigma_{\mathrm{max}}}$ as a function of $\ln{\rho}$ in figure \ref{difftimemaxdiffconst10}\subref{lndifftimespecdimmaxlndiffconst10}. The plot in figure \ref{difftimemaxdiffconst10}\subref{lndifftimespecdimmaxlndiffconst10} clearly exhibits scaling of $\sigma_{\mathrm{max}}$ with $\rho$. Fitting a line to $\ln{\sigma_{\mathrm{max}}}$ as a function of $\ln{\rho}$, I obtain a slope of $-0.99$. Evidently, $\langle \mathcal{D}_{\mathfrak{s}}(\sigma)\rangle$ depends trivially on $\rho$, matching the expectation from equation \eqref{heatequation}. Applying this scaling to $\langle \mathcal{D}_{\mathfrak{s}}(\sigma)\rangle$, I plot $\langle \mathcal{D}_{\mathfrak{s}}(\rho\sigma)\rangle$ in figure \ref{specdimscaleddiffconst}. 
The overlap of $\langle \mathcal{D}_{\mathfrak{s}}(\rho\sigma)\rangle$ for the five values of $\rho$ is clear. 

%Since I have established that $\langle\mathcal{D}_{\mathfrak{s}}(\sigma)\rangle$ exhibits trivial scaling with $\rho$, one might be tempted to compute $\langle\mathcal{D}_{\mathfrak{s}}(\sigma)\rangle$ for $\rho=1$ to reduce computational times. % or probe larger physical scales. 
%As Benedetti and Henson originally advocated \cite{}, and as figure \ref{} makes clear, a diffusion constant somewhat less than $1$ significantly smooths the spectral dimension for small $\sigma$, %While computing $\langle\mathcal{D}_{\mathfrak{s}}(\sigma)\rangle$ for $\rho=1$ makes for somewhat shorter computation times, quite possibly revealing the physical behavior of the spectral dimension for small $\sigma$ for smaller $N_{3}$. 
%If one's interests lie with the small-scale spectral dimension, then one should choose $\rho<1$, while, if one's interests lie with the large-scale spectral dimension, then one should choose $\rho=1$. 

\subsection{Dependence on the number $N_{3}$ of $3$-simplices}\label{N3dependence}

\begin{figure}[h]
\centering
\includegraphics[width=0.45\linewidth]{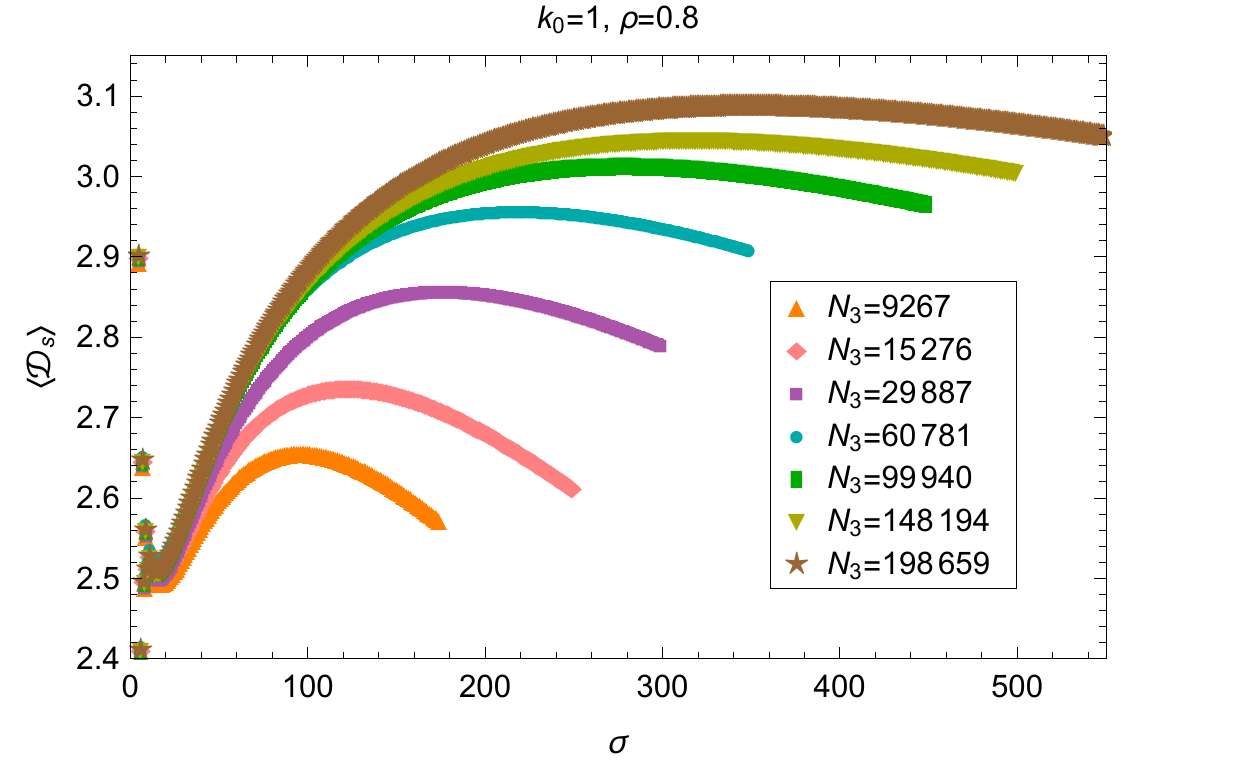}
\caption{The ensemble average spectral dimension $\langle\mathcal{D}_{\mathfrak{s}}\rangle$ as a function of the diffusion time $\sigma$ for seven values of the number $N_{3}$ of $3$-simplices (within the condensate) for $k_{0}=1.00$ and $\rho=0.8$.}
\label{specdimN3}
\end{figure}

I next explore the dependence of $\langle \mathcal{D}_{\mathfrak{s}}(\sigma)\rangle$ on $N_{3}$ holding fixed $k_{0}$ and $\rho$. I have performed measurements of $\langle \mathcal{D}_{\mathfrak{s}}(\sigma)\rangle$ for seven values of $N_{3}$ ranging from $1\times10^{5}$ to $2\times10^{6}$ at fixed $k_{0}$ and $\rho$ to determine its dependence on $N_{3}$. I display these measurements in figure \ref{specdimN3}.

\begin{figure}[h]
\centering
\includegraphics[width=0.45\linewidth]{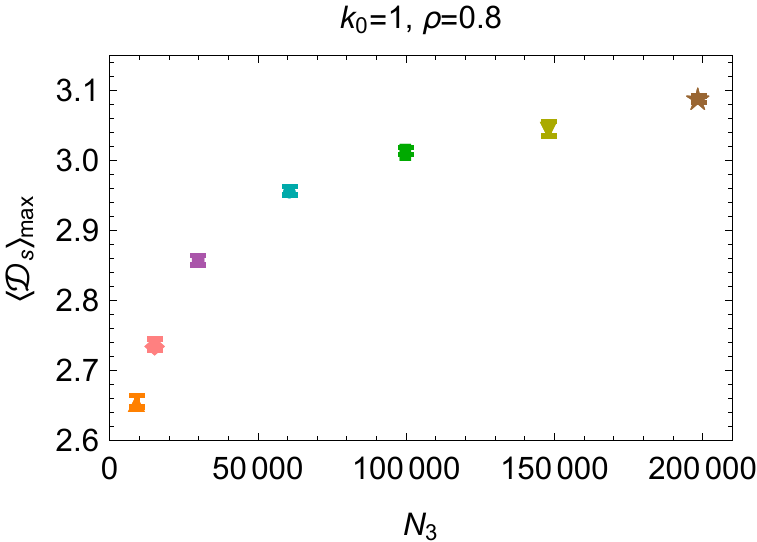}
\caption{The maximal value $\langle\mathcal{D}_{\mathfrak{s}}\rangle_{\mathrm{max}}$ of the ensemble average spectral dimension $\langle\mathcal{D}_{\mathfrak{s}}(\sigma)\rangle$ as a function of the number $N_{3}$ of $3$-simplices (within the condensate) for $k_{0}=1.00$ and $\rho=0.8$.}
\label{maxspecdimN3}
\end{figure}

\begin{figure}[h]
\centering
\subfigure[]{
\includegraphics[width=0.45\linewidth]{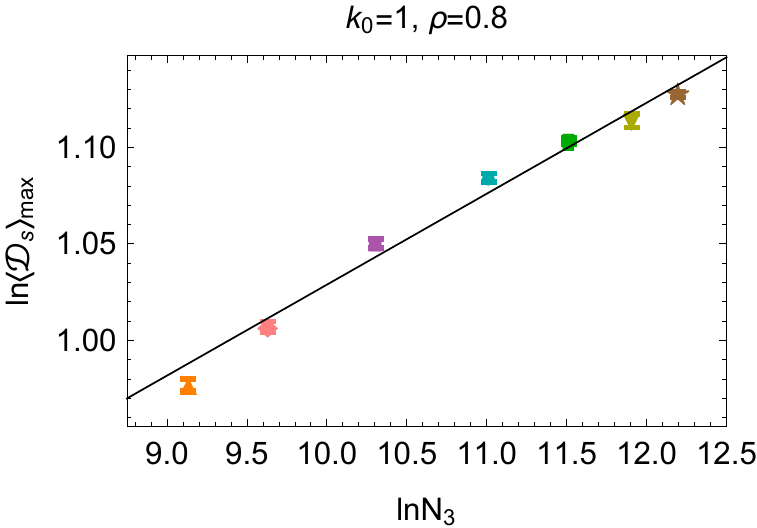}
\label{lnspecdimmaxlnN3fit}}
\subfigure[]{
\includegraphics[width=0.45\linewidth]{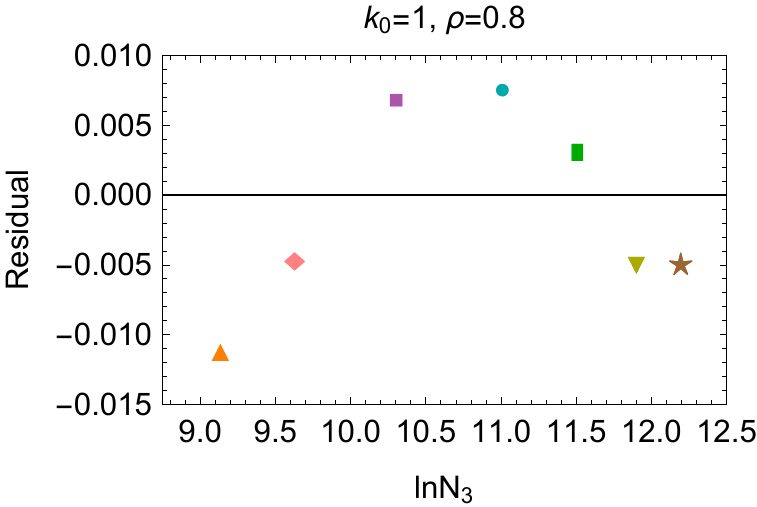}
\label{lnmaxspecdimlnN3res}}
\caption{\ref{lnmaxspecdimvslnN3}\subref{lnspecdimmaxlnN3fit} The logarithm $\ln{\langle\mathcal{D}_{\mathfrak{s}}\rangle_{\mathrm{max}}}$ of the maximal value $\langle\mathcal{D}_{\mathfrak{s}}\rangle_{\mathrm{max}}$ of the ensemble average spectral dimension $\langle\mathcal{D}_{\mathfrak{s}}(\sigma)\rangle$ as a function of the logarithm $\ln{N_{3}}$ of the number $N_{3}$ of $3$-simplices (within the condensate) for $k_{0}=1.00$ and $\rho=0.8$ overlain with the linear least-squares fit $\ln{\langle\mathcal{D}_{\mathfrak{s}}\rangle_{\mathrm{max}}}=0.56+0.05\ln{N_{3}}$. \ref{lnmaxspecdimvslnN3}\subref{lnmaxspecdimlnN3res} The residuals of the linear least-squares fit $\ln{\langle\mathcal{D}_{\mathfrak{s}}\rangle_{\mathrm{max}}}=0.56+0.05\ln{N_{3}}$ to the logarithm $\ln{\langle\mathcal{D}_{\mathfrak{s}}\rangle_{\mathrm{max}}}$ of the maximal value $\langle\mathcal{D}_{\mathfrak{s}}\rangle_{\mathrm{max}}$ of the ensemble average spectral dimension $\langle\mathcal{D}_{\mathfrak{s}}(\sigma)\rangle$ as a function of the logarithm $\ln{N_{3}}$ of the number $N_{3}$ of $3$-simplices (within the condensate) for $k_{0}=1.00$ and $\rho=0.8$.}
\label{lnmaxspecdimvslnN3}
\end{figure}

I again focus on the maximum of $\langle \mathcal{D}_{\mathfrak{s}}(\sigma)\rangle$ as a clearly identifiable point of comparison. I first plot $\langle \mathcal{D}_{\mathfrak{s}}\rangle_{\mathrm{max}}$ as a function of $N_{3}$ in figure \ref{maxspecdimN3}.
Previous studies indicate that $\langle\mathcal{D}_{\mathfrak{s}}\rangle_{\mathrm{max}}$ approaches the topological dimension of $3$ from below in the infinite volume limit, possibly slightly overshooting \cite{DB&JH}.\footnote{$\langle\mathcal{D}_{\mathfrak{s}}\rangle_{\mathrm{max}}$ approaches the topological dimension of $4$ from below in the infinite volume limit within phase C of the causal dynamical triangulations of $(3+1)$-dimensional Einstein gravity \cite{JA&JJ&RL7,JA&JJ&RL6,DNC&JJ}.} The measurements of $\langle \mathcal{D}_{\mathfrak{s}}\rangle_{\mathrm{max}}$ in figure \ref{maxspecdimN3} corroborate these studies: $\langle\mathcal{D}_{\mathfrak{s}}\rangle_{\mathrm{max}}$ approaches and slightly overshoots $3$ as $N_{3}$ increases. Indeed, a value of $3$ for $\langle\mathcal{D}_{\mathfrak{s}}\rangle_{\mathrm{max}}$ does not fall within the range of the measurement errors for the three largest values of $N_{3}$. The plot in figure \ref{maxspecdimN3} also indicates that $\langle\mathcal{D}_{\mathfrak{s}}\rangle_{\mathrm{max}}$ continues to increase for $N_{3}>2\times10^{6}$, so the extent of the overshoot in the infinite volume limit may well be more than $0.08$. 
%Using the measurements presented in figure \ref{maxspecdimN3}, 
I wish to determine if $\langle\mathcal{D}_{\mathfrak{s}}\rangle_{\mathrm{max}}$ is at least finite in the infinite volume limit. % and extrapolate the overshoot in this limit. I wish to extrapolate $\langle\mathcal{D}_{\mathfrak{s}}\rangle_{\mathrm{max}}$ towards the infinite volume limit. Is $\langle\mathcal{D}_{\mathfrak{s}}\rangle_{\mathrm{max}}$ finite in the infinite volume limit? If $\langle\mathcal{D}_{\mathfrak{s}}\rangle_{\mathrm{max}}$ is finite in the infinite volume limit, then is its value precisely $3$? 
In light of the regular increase of $\langle\mathcal{D}_{\mathfrak{s}}\rangle_{\mathrm{max}}$ with $N_{3}$ exhibited in figure \ref{maxspecdimN3}, I test for scaling of $\langle\mathcal{D}_{\mathfrak{s}}\rangle_{\mathrm{max}}$ with $N_{3}$ by plotting $\ln{\langle\mathcal{D}_{\mathfrak{s}}\rangle_{\mathrm{max}}}$ as a function of $\ln{N_{3}}$ in figure \ref{lnmaxspecdimvslnN3}\subref{lnspecdimmaxlnN3fit}. 
Performing a linear least squares fit of $\ln{\langle\mathcal{D}_{\mathfrak{s}}\rangle_{\mathrm{max}}}$ as a function of $\ln{N_{3}}$, shown as a black line in figure \ref{lnmaxspecdimvslnN3}\subref{lnspecdimmaxlnN3fit}, I find that
\begin{equation}
\ln{\langle\mathcal{D}_{\mathfrak{s}}\rangle_{\mathrm{max}}}=0.56+0.05\ln{N_{3}},
\end{equation}
implying that
\begin{equation}
\langle\mathcal{D}_{\mathfrak{s}}\rangle_{\mathrm{max}}(N_{3})\propto N_{3}^{0.05}.
\end{equation}
I plot the residuals of this fit as a function of $\ln{N_{3}}$ in figure \ref{lnmaxspecdimvslnN3}\subref{lnmaxspecdimlnN3res}. These residuals clearly show a trend, indicating that $\langle\mathcal{D}_{\mathfrak{s}}\rangle_{\mathrm{max}}$ does not increase as a power of $N_{3}$ but rather increases as a subpower of $N_{3}$. %Accounting for measurement errors, the scaling of $\mathcal{D}_{s}^{\mathrm{max}}$ with $N_{3}$ is also consistent with a slightly negative power. I require further measurements to determine definitively the scaling of $\mathcal{D}_{s}^{\mathrm{max}}$ with $N_{3}$. 
Since a function $f(x)$ that increases as a subpower of $x$ is finite in the limit $x\rightarrow\infty$, I conclude that $\langle\mathcal{D}_{\mathfrak{s}}\rangle_{\mathrm{max}}$ is finite in the infinite volume limit. Moreover, the findings of subsection \ref{k0dependence} below admit the possibility of making $\langle\mathcal{D}_{\mathfrak{s}}\rangle_{\mathrm{max}}$ equal to $3$ in the infinite volume limit.

%I analyze two aspects of these measurements: %of $\langle \mathcal{D}_{\mathfrak{s}}(\sigma)\rangle$ displayed in figure \ref{specdimN3}: 
%the dependence of $\langle\mathcal{D}_{\mathfrak{s}}\rangle_{\mathrm{max}}$ on $N_{3}$ and the dependence of $\sigma_{\mathrm{max}}$ on $N_{3}$. Specifically, I wish to address the following three questions. %I with to address three questions through this analysis, two regarding $\langle\mathcal{D}_{\mathfrak{s}}\rangle_{\mathrm{max}}$ and one regarding $\sigma_{\mathrm{max}}$. 
%Is $\langle\mathcal{D}_{\mathfrak{s}}\rangle_{\mathrm{max}}$ finite in the infinite volume limit? If $\langle\mathcal{D}_{\mathfrak{s}}\rangle_{\mathrm{max}}$ is finite in the infinite volume limit, then is its value precisely $3$? How does $\sigma_{\mathrm{max}}$ scale with $N_{3}$?

%The scaling of the $\sigma$-dependence of $\langle \mathcal{D}_{\mathfrak{s}}(\sigma)\rangle$ with $N_{3}$ has not previously been investigated. I perform measurements of $\langle \mathcal{D}_{\mathfrak{s}}(\sigma)\rangle$ to determine its dependence on $N_{3}$. 

\begin{figure}[h]
\centering
\includegraphics[width=0.45\linewidth]{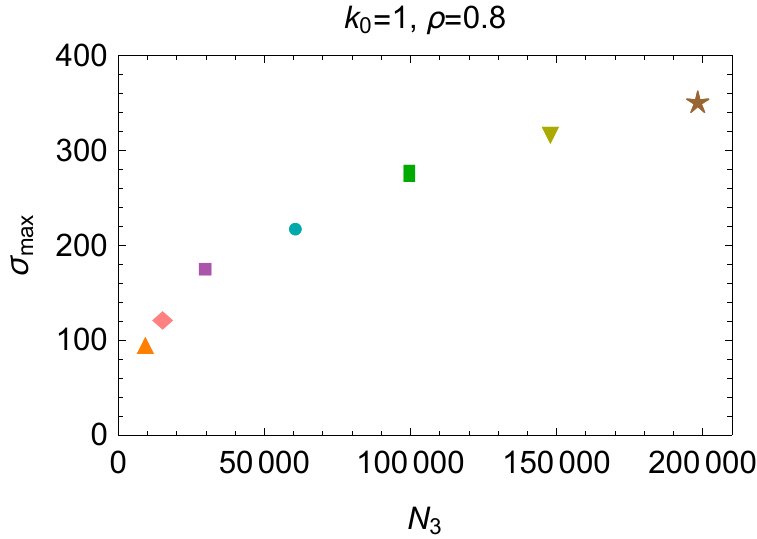}
\caption{The diffusion time $\sigma_{\mathrm{max}}$ as a function of the number $N_{3}$ of $3$-simplices (within the condensate) for $k_{0}=1.00$ and $\rho=0.8$.}
\label{sigmamaxvsN3}
\end{figure}

\begin{figure}[h]
\centering
\subfigure[]{
\includegraphics[width=0.45\linewidth]{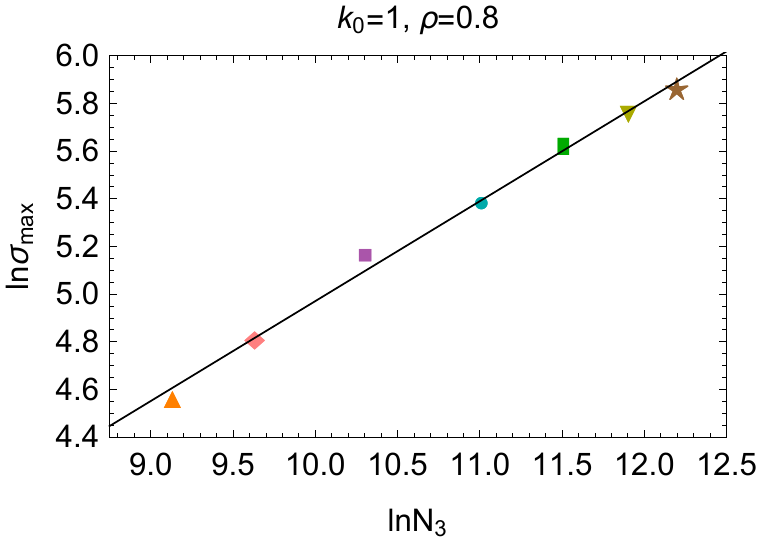}
\label{difftimespecdimmaxN3}}
\subfigure[]{
\includegraphics[width=0.45\linewidth]{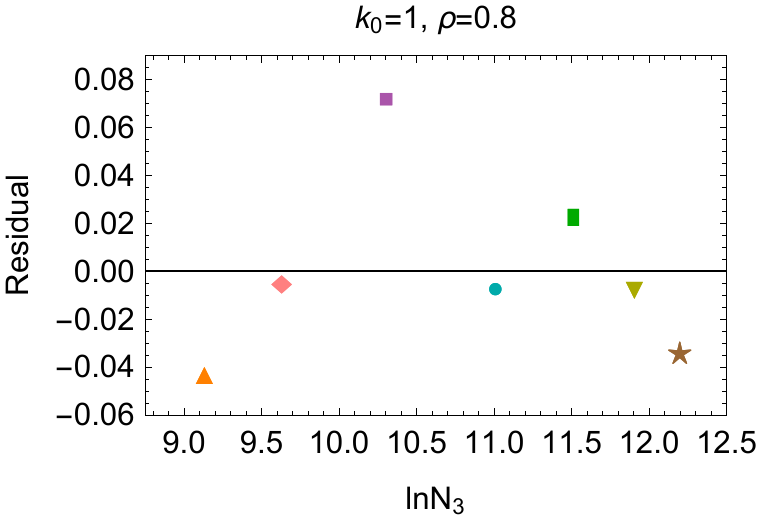}
\label{lndifftimespecdimmaxlnN3}}
\caption{\subref{difftimespecdimmaxN3} The logarithm $\ln{\sigma_{\mathrm{max}}}$ of the diffusion time $\sigma_{\mathrm{max}}$ as a function of the logarithm $\ln{N_{3}}$ of the number $N_{3}$ of $3$-simplices (within the condensate) for $k_{0}=1.00$ and $\rho=0.8$ overlain in black with the linear least squares fit $\ln{\sigma_{\mathrm{max}}}=0.78+0.42\ln{N_{3}}$. \subref{lndifftimespecdimmaxlnN3} The residuals of the linear least squares fit of the logarithm $\ln{\sigma_{\mathrm{max}}}$ of the diffusion time $\sigma_{\mathrm{max}}$ as a function of the logarithm $\ln{N_{3}}$ of the number $N_{3}$ of $3$-simplices (within the condensate) for $k_{0}=1.00$ and $\rho=0.8$.}
\label{difftimespecdimvsmaxN3}
\end{figure}

I next plot $\sigma_{\mathrm{max}}$ as a function of $N_{3}$ in figure \ref{sigmamaxvsN3}.
%To determine the scaling of the $\sigma$-dependence of $\langle \mathcal{D}_{s}(\sigma)\rangle$ with $N_{3}$, 
In light of the regular increase of $\sigma_{\mathrm{max}}$ with $N_{3}$ exhibited in figure \ref{sigmamaxvsN3}, I test for scaling of $\sigma_{\mathrm{max}}$ with $N_{3}$ by plotting $\ln{\sigma_{\mathrm{max}}}$ as a function of $\ln{N_{3}}$ in figure \ref{difftimespecdimvsmaxN3}\subref{difftimespecdimmaxN3}.
The plot in figure \ref{difftimespecdimvsmaxN3}\subref{difftimespecdimmaxN3} shows evidence of scaling of $\sigma_{\mathrm{max}}$ with $N_{3}$: there is an approximately linear relationship between $\ln{\sigma_{\mathrm{max}}}$ and $\ln{N_{3}}$. Performing a linear least squares fit of $\ln{\sigma_{\mathrm{max}}}$ as a function of $\ln{N_{3}}$, shown as a black line in figure \ref{difftimespecdimvsmaxN3}\subref{difftimespecdimmaxN3}, I find that 
\begin{equation}
\ln{\sigma_{\mathrm{max}}}=0.78+0.42\ln{N_{3}}, 
\end{equation}
implying that
\begin{equation}\label{measuredsigmamaxscaling}
\sigma_{\mathrm{max}}(N_{3})\propto N_{3}^{0.42}.
\end{equation}
I plot the residuals of this fit as a function of $\ln{N_{3}}$ in figure \ref{difftimespecdimvsmaxN3}\subref{lndifftimespecdimmaxlnN3}. Excluding the residual for $N_{3}=60781$ (dark cyan, circle), the residuals show a trend, possibly indicating that $\sigma_{\mathrm{max}}$ does not increase as a power of $N_{3}$ but rather as a subpower of $N_{3}$. %Do we expect $\sigma_{\mathrm{max}}$ to grow continuously with $N_{3}$?

Previous studies possibly suggest that $\sigma_{\mathrm{max}}$ diverges in the infinite volume limit. These studies have advanced evidence for and made use of the double scaling limit
\begin{equation}\label{doublescalinglimit}
V_{3}=\lim_{\substack{N_{3}\rightarrow\infty \\ a\rightarrow0}}C_{3}N_{3}a^{3}
\end{equation}
for the spacetime $3$-volume $V_{3}$ \cite{JA&JJ&RL3,DB&JH,DB&JH2,JHC&KL&JMM,JHC&JMM}: in the combination of the infinite volume limit ($N_{3}\rightarrow\infty$) and the continuum limit ($a\rightarrow0$), $C_{3}N_{3}a^{3}$ approaches $V_{3}$. $C_{3}$ is the effective discrete spacetime $3$-volume of a single $3$-simplex, so $N_{3}$ is essentially the discrete spacetime $3$-volume. Assuming that $V_{3}$ is fixed (for fixed $k_{0}$) and that the double scaling limit \eqref{doublescalinglimit} holds with negligible corrections for finite but sufficiently large $N_{3}$ and finite but sufficiently small $a$, one concludes that $a$ scales with $N_{3}$ as $N_{3}^{-1/3}$. Accordingly, a discrete quantity associated with units of $a^{p}$ for some power $p$ (canonically) scales with $N_{3}$ as $N_{3}^{-p/3}$. %Applying this reasoning to $\sigma$, 
The diffusion time $\sigma$ is associated with units of $a^{2}$ since %as justified by the analysis of subsection \ref{diffconstdependence}, 
the diffusion time $s$ has dimensions of squared length (after absorbing the diffusion constant $\wp$ into $s$ in equation \eqref{heatequation}). On the basis of the double scaling limit \eqref{doublescalinglimit}, one therefore expects $\sigma$ to scale with $N_{3}$ as $N_{3}^{-2/3}$. %to be proportional to $s$ in the continuum limit. % That $N_{3}$ is essentially the discrete spacetime $3$-volume 
Denoting by $s_{\mathrm{max}}$ the fixed value of $s$ at which $D_{\mathfrak{s}}(s)$ attains its maximum, one infers that
\begin{equation}\label{smaxscaling}
s_{\mathrm{max}}\propto\sigma_{\mathrm{max}}(N_{3})\,a^{2}.
\end{equation}
The double scaling limit \eqref{doublescalinglimit} and relation \eqref{smaxscaling} then dictate that
\begin{equation}\label{sigmamaxscaling}
\sigma_{\mathrm{max}}(N_{3})\propto N_{3}^{2/3},
\end{equation}
implying that $\sigma_{\mathrm{max}}$ diverges in the infinite volume limit. 

The evidence for the double scaling limit \eqref{doublescalinglimit} stems from analysis of large-scale observables \cite{JA&JJ&RL6}, and the uses of the double scaling limit \eqref{doublescalinglimit} come in modeling of large-scale observables \cite{CA&SC&JHC&PH&RKK&PZ,DB&JH,DB&JH2,JHC&KL&JMM,JHC&JMM}. For instance, Ambj\o rn, Jurkiewicz, and Loll originally arrived at the double scaling limit \eqref{doublescalinglimit} by studying temporal correlations in the discrete spatial volume \cite{JA&JJ&RL6}, and Benedetti and Henson applied the double scaling limit \eqref{doublescalinglimit} to the diffusion time in fitting the spectral dimension of deformed de Sitter space to the large-scale ensemble average spectral dimension \cite{DB&JH}. %They apply this canonical scaling to values of $\sigma$ considerably larger than $\sigma_{\mathrm{max}}$. 
Accordingly, I only expect $\sigma$ to scale with $N_{3}$ as $N_{3}^{-2/3}$ %, leading to the relation \eqref{sigmamaxscaling}, 
for sufficiently large values of $\sigma$, presumably for random walks that probe classical or at least semiclassical regimes. 

Evidently, $\sigma_{\mathrm{max}}$ does not follow the relation \eqref{sigmamaxscaling} but scales noncanonically with $N_{3}$. Presumably, for $\sigma<\sigma_{\mathrm{max}}$, $\sigma$ also scales noncanonically with $N_{3}$. While one should have anticipated noncanonical scaling of $\sigma$ with $N_{3}$ for $\sigma<\sigma_{\mathrm{max}}$ since $\langle \mathcal{D}_{\mathfrak{s}}(\sigma)\rangle$ exhibits the quantum-mechanical effect of dynamical dimensional reduction for $\sigma<\sigma_{\mathrm{max}}$, one might have expected $\sigma_{\mathrm{max}}$ to scale canonically with $N_{3}$ since $\langle \mathcal{D}_{\mathfrak{s}}(\sigma)\rangle$ attains the topological dimension of $3$ on the scale associated with $\sigma_{\mathrm{max}}$. This finding is consistent with the analysis of \cite{DB&JH} in which canonical scaling of $\sigma$ with $N_{3}$ was only applied for $\sigma>\sigma_{\mathrm{max}}$. %consistent with $\frac{1}{2}$ given the errors.
%\begin{figure}[h]
%\includegraphics[width=\linewidth]{specdimscaledN3.pdf}
%\end{figure}
The noncanonical scaling of $\sigma_{\mathrm{max}}$ with $N_{3}$ expressed in equation \eqref{measuredsigmamaxscaling} suggests that the quantum geometry on the scale associated with $\sigma_{\mathrm{max}}$ is not as semiclassical as the value of $\langle \mathcal{D}_{\mathfrak{s}}(\sigma)\rangle$ suggests.

Numerical measurements of $\langle\mathcal{D}_{\mathfrak{s}}(\sigma)\rangle$ for $N_{3}>2\times10^{5}$ should determine whether $\sigma_{\mathrm{max}}$ is finite or infinite in the infinite volume limit. Despite the inconclusiveness of the above analysis, $\sigma_{\mathrm{max}}$ presumably diverges in the infinite volume limit taken alone: as $N_{3}$ diverges, $V_{3}$ diverges, and, as $V_{3}$ defines the largest scale characterizing a causal triangulation, any smaller scales can likewise diverge. Conversely, $\sigma_{\mathrm{max}}$ presumably remains finite in the double scaling limit \eqref{doublescalinglimit}, the infinite volume and continuum limits taken together: as $N_{3}$ diverges and $a$ vanishes, $V_{3}$ remains finite, and, as $V_{3}$ defines the largest scale characterizing a causal triangulation, any smaller scales must likewise remain finite.

\subsection{Dependence on the bare coupling $k_{0}$}\label{k0dependence}

\begin{figure}[h]
\centering
\includegraphics[width=\linewidth]{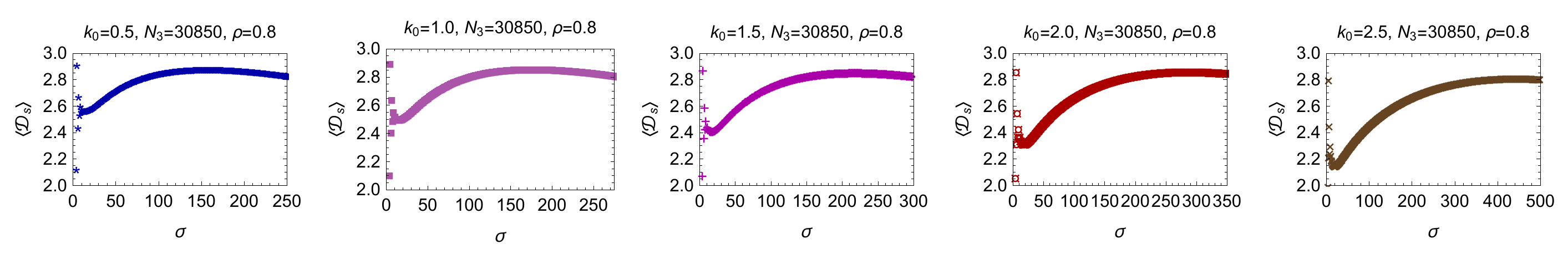}
\caption{The ensemble average spectral dimension $\langle\mathcal{D}_{\mathfrak{s}}\rangle$ as a function of the diffusion time $\sigma$ for bare couplings $k_{0}=0.5$, $k_{0}=1.0$, $k_{0}=1.5$, $k_{0}=2.0$, and $k_{0}=2.5$ for $N_{3}=30850$, $T=64$, and $\rho=0.8$.}
\label{specdimvaryingk0sep}
\end{figure}

\begin{figure}[h]
\centering
\includegraphics[width=0.45\linewidth]{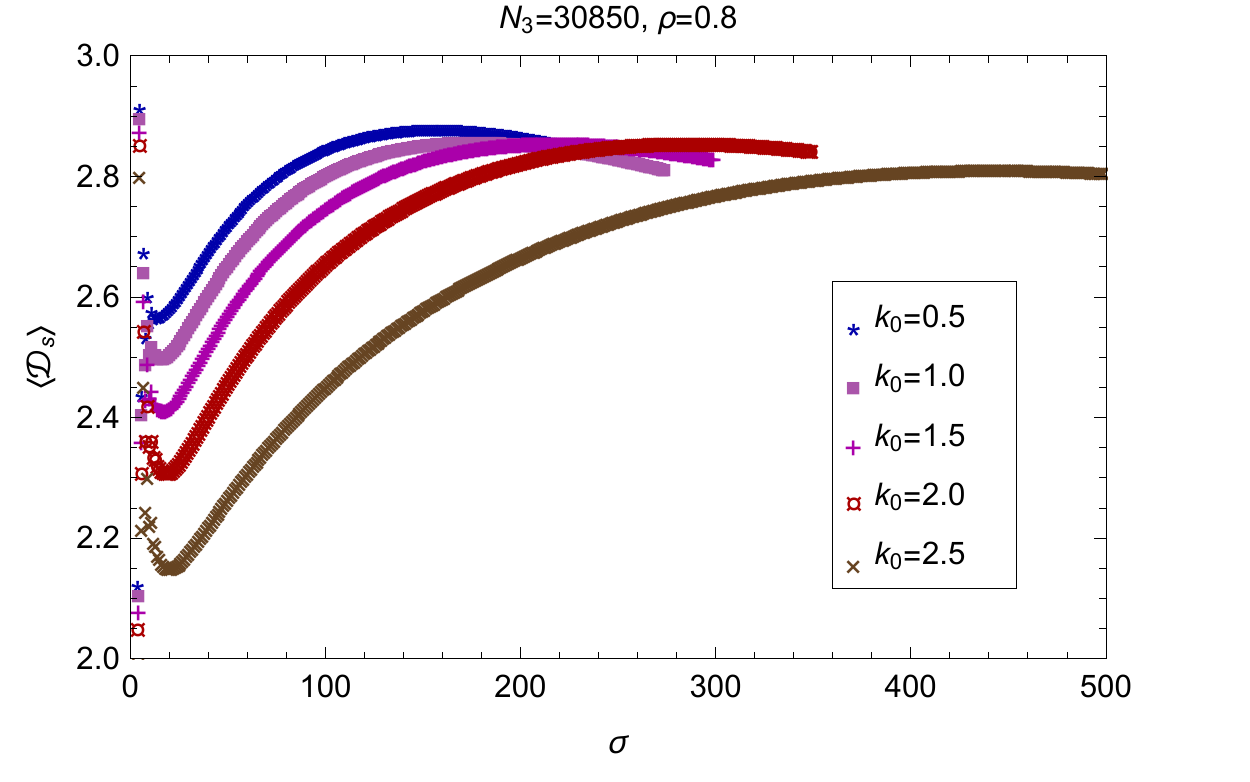}
\caption{The ensemble average spectral dimension $\langle\mathcal{D}_{\mathfrak{s}}\rangle$ as a function of the diffusion time $\sigma$ for five values of the bare coupling $k_{0}$ for $N_{3}=30850$, $T=64$, and $\rho=0.8$.}
\label{specdimvaryingk0}
\end{figure}

I finally explore the dependence of $\langle\mathcal{D}_{\mathfrak{s}}(\sigma)\rangle$ on $k_{0}$ holding fixed $\rho$ and $N_{3}$. I have performed measurements of $\langle\mathcal{D}_{\mathfrak{s}}(\sigma)\rangle$ for five values of $k_{0}$ to determine its dependence on $k_{0}$. I display these measurements separately in figure \ref{specdimvaryingk0sep} and together in figure \ref{specdimvaryingk0}. 
The plots in figures \ref{specdimvaryingk0sep} and \ref{specdimvaryingk0} show that $\langle\mathcal{D}_{\mathfrak{s}}(\sigma)\rangle$ changes with $k_{0}$ just as $\langle\mathcal{D}_{\mathfrak{s}}(\sigma)\rangle$ changes with the corresponding bare coupling $\kappa_{0}$ within phase C of the causal dynamical triangulations of $4$-dimensional Einstein gravity \cite{DNC&JJ}. This constitutes another respect in which the phenomenology of the $3$-dimensional model reproduces that of the $4$-dimensional model. 

\begin{figure}[h]
\centering
\includegraphics[width=0.45\linewidth]{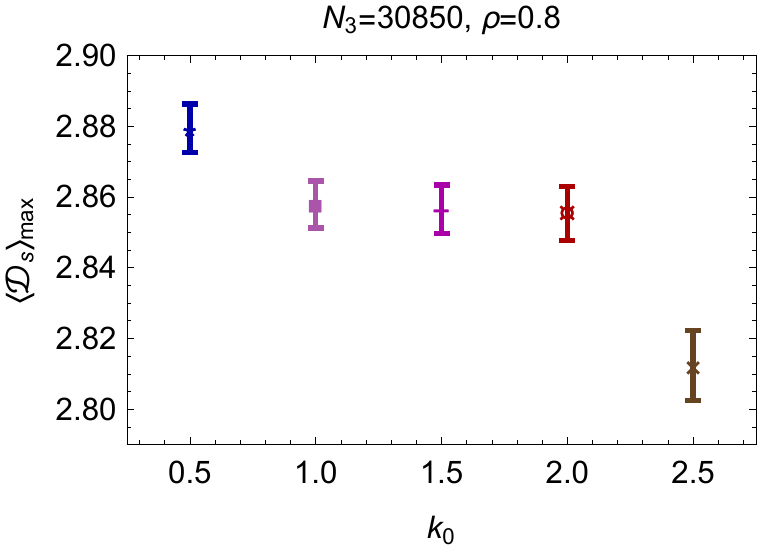}
\caption{The maximal value $\langle\mathcal{D}_{\mathfrak{s}}\rangle_{\mathrm{max}}$ of the ensemble average spectral dimension $\langle\mathcal{D}_{\mathfrak{s}}(\sigma)\rangle$ as a function of the bare coupling $k_{0}$ for $N_{3}=30850$, $T=64$, and $\rho=0.8$.}
\label{specdimmaxvaryingk0}
\end{figure}

I once again focus on the maximum of $\langle\mathcal{D}_{\mathfrak{s}}(\sigma)\rangle$ as a clearly identifiable point of comparison. I first plot $\langle\mathcal{D}_{\mathfrak{s}}\rangle_{\mathrm{max}}$ as a function of $k_{0}$ in figure \ref{specdimmaxvaryingk0}. 
The plot in figure \ref{specdimmaxvaryingk0} shows that $\langle\mathcal{D}_{\mathfrak{s}}\rangle_{\mathrm{max}}$ is a constant function of $k_{0}$ (up to measurement errors) for $k_{0}$ well within phase C, that $\langle\mathcal{D}_{\mathfrak{s}}\rangle_{\mathrm{max}}$ increases as $k_{0}$ approaches $0$, and that $\langle\mathcal{D}_{\mathfrak{s}}\rangle_{\mathrm{max}}$ decreases as $k_{0}$ approaches the AC phase transition. Note, though, that, ignoring measurement errors, $\langle\mathcal{D}_{\mathfrak{s}}\rangle_{\mathrm{max}}$ decreases monotonically with $k_{0}$, albeit very slowly well within phase C. One might suspect that this decrease of $\langle\mathcal{D}_{\mathfrak{s}}\rangle_{\mathrm{max}}$ with $k_{0}$ is merely a finite-size effect since the number of $3$-simplices within the condensate also decreases with $k_{0}$. (See figure \ref{} below.) An analysis of the number of $3$-simplices within the condensate as a function of $k_{0}$ indicates that this suspicion is unsubstantiated (unless proximity to the AC phase transition amplifies the finite-size effect). % at $k_{0}\approx3.3$. 

\begin{figure}[h]
\centering
\subfigure[]{
\includegraphics[width=0.45\linewidth]{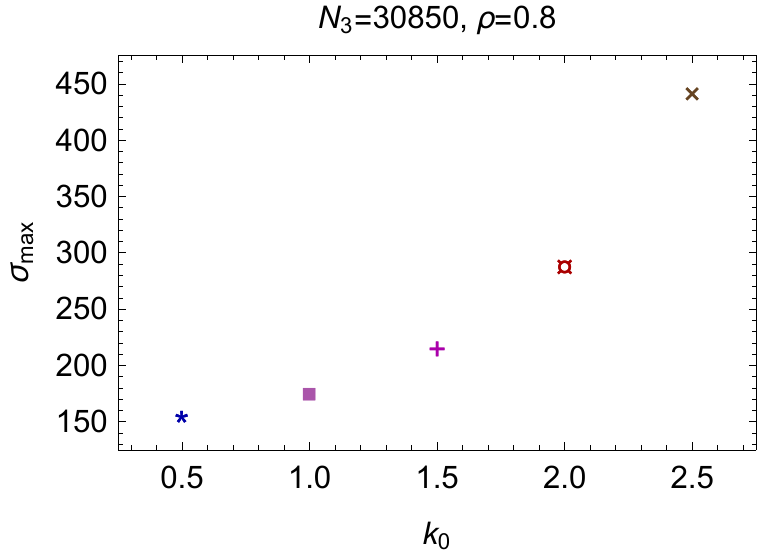}
\label{sigmamaxk0}}
\subfigure[]{
\includegraphics[width=0.45\linewidth]{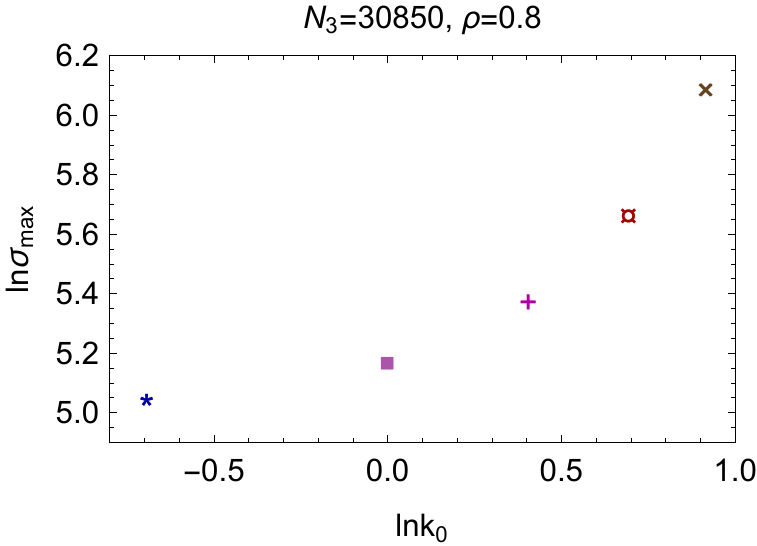}
\label{lnsigmamaxlnk0}}
\caption{\subref{sigmamaxk0} The diffusion time $\sigma_{\mathrm{max}}$ as a function of  the bare coupling $k_{0}$ for $N_{3}=30850$, $T=64$, and $\rho=0.8$. \subref{lnsigmamaxlnk0} The logarithm $\ln{\sigma_{\mathrm{max}}}$ of the diffusion time $\sigma_{\mathrm{max}}$ as a function of the logarithm $\ln{k_{0}}$ of the bare coupling $k_{0}$ for $N_{3}=30850$, $T=64$, and $\rho=0.8$.}
\label{sigmamaxvaryingk0}
\end{figure}

I next plot $\sigma_{\mathrm{max}}$ as a function of $k_{0}$ in figure \ref{sigmamaxvaryingk0}\subref{sigmamaxk0}. 
In light of the regular increase of $\sigma_{\mathrm{max}}$ with $k_{0}$ exhibited in figure \ref{sigmamaxvaryingk0}\subref{sigmamaxk0}, I test for scaling of $\sigma_{\mathrm{max}}$ with $k_{0}$ by plotting $\ln{\sigma_{\mathrm{max}}}$ as a function of $\ln{k_{0}}$ in figure \ref{sigmamaxvaryingk0}\subref{lnsigmamaxlnk0}. The plot in figure \ref{sigmamaxvaryingk0}\subref{lnsigmamaxlnk0} shows that $\sigma_{\mathrm{max}}$ increases more rapidly with $k_{0}$ than as a power of $k_{0}$; indeed, I conjecture in section \ref{conclusion} that $\sigma_{\mathrm{max}}$ diverges as $k_{0}$ approaches the AC phase transition. %with phase $A$ at $k_{0}\approx3.3$. 

\begin{figure}[h]
\centering
\subfigure[]{
\includegraphics[width=0.45\linewidth]{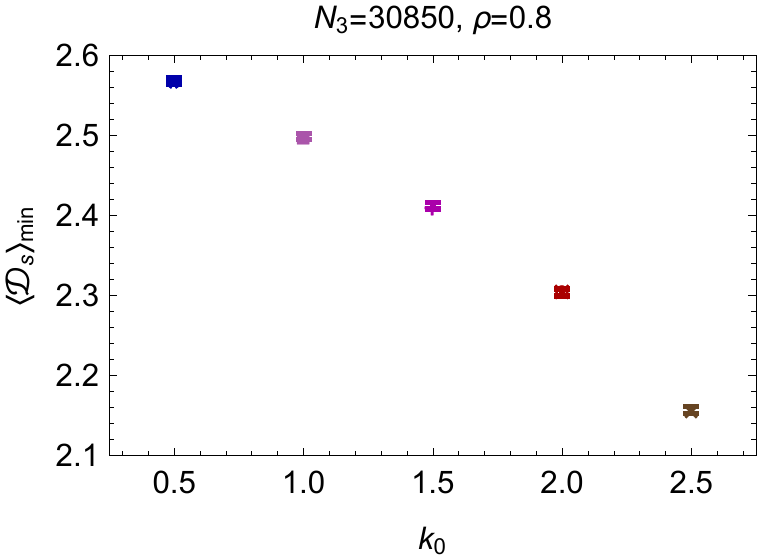}
\label{specdimmink0}}
\subfigure[]{
\includegraphics[width=0.45\linewidth]{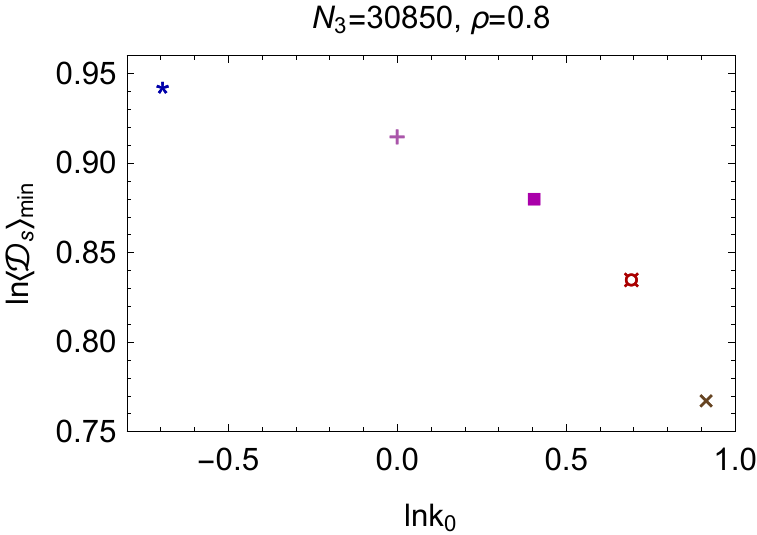}
\label{lnspecdimminlnk0}}
\caption{\subref{specdimmink0} The minimal value $\langle\mathcal{D}_{\mathfrak{s}}\rangle_{\mathrm{min}}$ of the ensemble average spectral dimension $\langle\mathcal{D}_{\mathfrak{s}}\rangle$ as a function of the bare coupling $k_{0}$ for $N_{3}=30850$, $T=64$, and $\rho=0.8$. \subref{lnspecdimminlnk0} The logarithm $\ln{\langle\mathcal{D}_{\mathfrak{s}}\rangle_{\mathrm{min}}}$ of the minimal value $\langle\mathcal{D}_{\mathfrak{s}}\rangle_{\mathrm{min}}$ of the ensemble average spectral dimension $\langle\mathcal{D}_{\mathfrak{s}}\rangle$ as a function of the logarithm $\ln{k_{0}}$ of the bare coupling $k_{0}$ for $N_{3}=30850$, $T=64$, and $\rho=0.8$.}
\label{specdimminvaryingk0}
\end{figure}

I also consider the minimum of $\langle\mathcal{D}_{\mathfrak{s}}(\sigma)\rangle$.\footnote{I exclude the (presumably) unphysically small values of $\langle\mathcal{D}_{\mathfrak{s}}(\sigma)\rangle$ for very small values of $\sigma$ ($\sigma<15$ for $k_{0}=0.5$, $\sigma<9$ for $k_{0}=1.0$, $\sigma<7$ for $k_{0}=1.5$, $\sigma<5$ for $k_{0}=2.0$, $\sigma<5$ for $k_{0}=2.5$).} I plot $\langle\mathcal{D}_{\mathfrak{s}}\rangle_{\mathrm{min}}$ as a function of $k_{0}$ in figure \ref{specdimminvaryingk0}\subref{specdimmink0}. 
In light of the regular decrease of $\langle\mathcal{D}_{\mathfrak{s}}\rangle_{\mathrm{min}}$ with $k_{0}$, I test for scaling of $\langle\mathcal{D}_{\mathfrak{s}}\rangle_{\mathrm{min}}$ with $k_{0}$ by plotting $\ln{\langle\mathcal{D}_{\mathfrak{s}}\rangle_{\mathrm{min}}}$ as a function of $\ln{k_{0}}$ in figure \ref{specdimminvaryingk0}\subref{lnspecdimminlnk0}. The plot in figure \ref{specdimminvaryingk0}\subref{lnspecdimminlnk0} shows that $\langle\mathcal{D}_{\mathfrak{s}}\rangle_{\mathrm{min}}$ decreases more rapidly with $k_{0}$ than as a power of $k_{0}$; nevertheless, I conjecture in section \ref{conclusion} that $\langle\mathcal{D}_{\mathfrak{s}}\rangle_{\mathrm{min}}$ approaches a nonzero value at the AC phase transition. 

\begin{figure}[h]
\centering
\subfigure[]{
\includegraphics[width=0.45\linewidth]{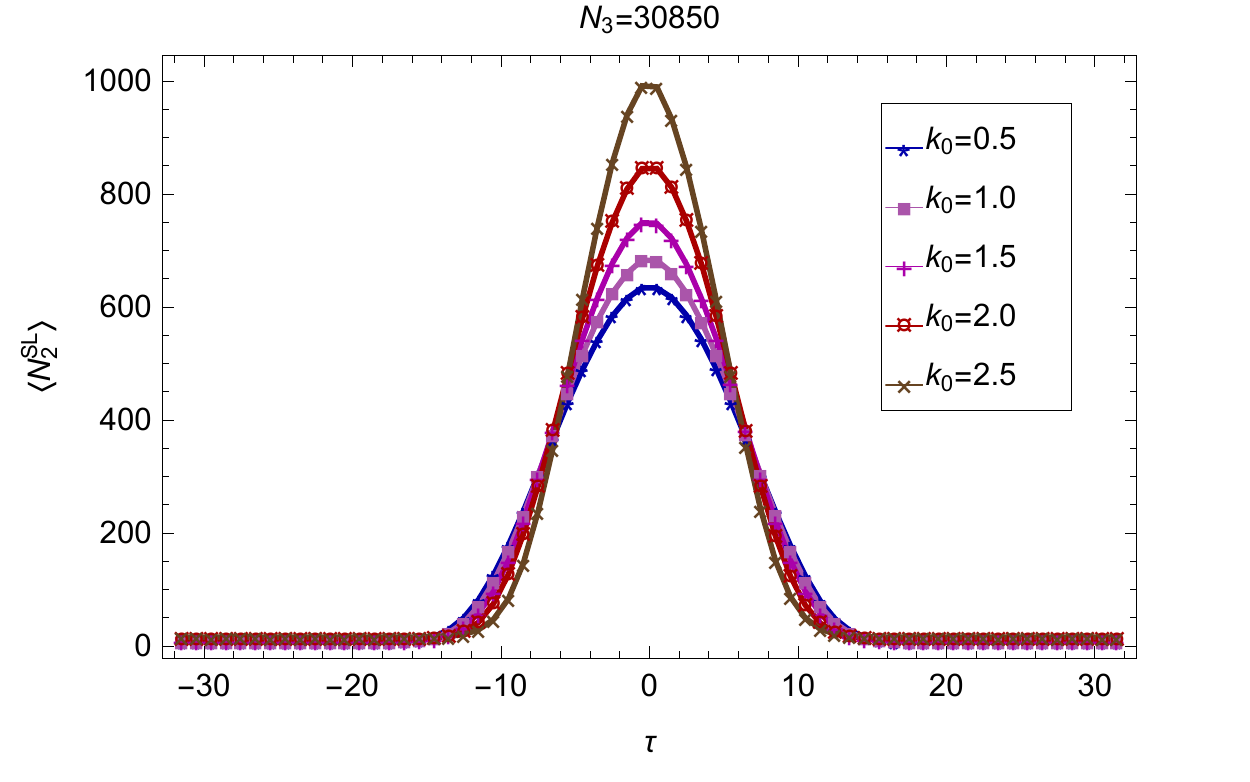}
\label{volprofilek0}}
\subfigure[]{
\includegraphics[width=0.45\linewidth]{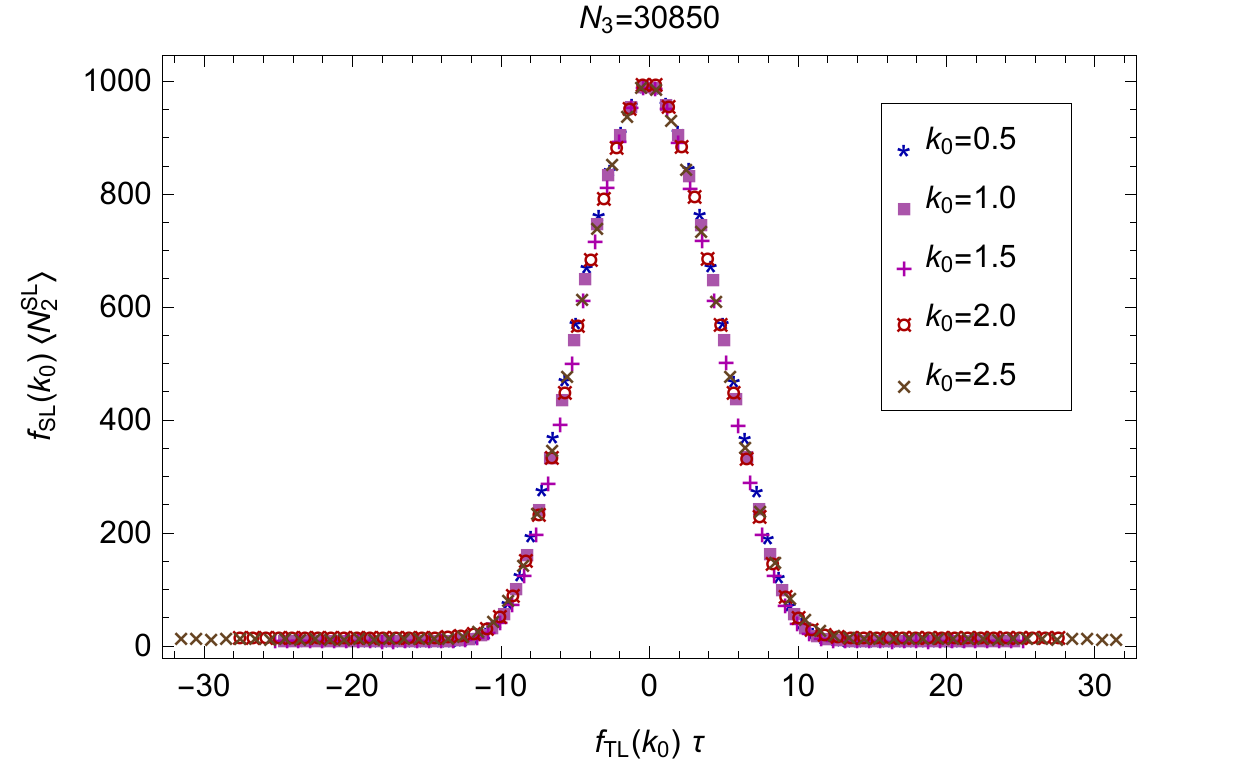}
\label{scaledvolprofilek0}}
\caption{\subref{volprofilek0} The ensemble average number $\langle N_{2}^{\mathrm{SL}}\rangle$ of spacelike $2$-simplices as a function of the discrete time coordinate $\tau$ for five values of the bare coupling $k_{0}$ for $N_{3}=30850$ and $T=64$. \subref{scaledvolprofilek0} The ensemble average number $f_{\mathrm{SL}}(k_{0})\,\langle N_{2}^{\mathrm{SL}}\rangle$ of spacelike $2$-simplices, scaled by the function $f_{\mathrm{SL}}(k_{0})$, as a function of the discrete time coordinate $f_{\mathrm{TL}}(k_{0})\,\tau$, scaled by the function $f_{\mathrm{TL}}(k_{0})$, for five values of the bare coupling $k_{0}$ for $N_{3}=30850$ and $T=64$.}
\label{volprofile}
\end{figure}

Ambj\o rn, Jurkiewicz, and Loll conjectured that $k_{0}$ only sets the overall scale of the quantum geometry within phase C \cite{JA&JJ&RL3}. Numerical measurements of the temporal evolution of the spatial $2$-volume---specifically, the ensemble average number $\langle N_{2}^{\mathrm{SL}}(\tau)\rangle$ of spacelike $2$-simplices as a function of the discrete time coordinate $\tau$ labeling the $T$ leaves of the global foliation---constituted their primary evidence. These authors found that, for different values of $k_{0}$, $\langle N_{2}^{\mathrm{SL}}(\tau)\rangle$ differs only by rescaling by a function of $k_{0}$. %, indicating that these different measurements encoded the same information. 
I have performed measurements of $\langle N_{2}^{\mathrm{SL}}(\tau)\rangle$ for the five values of $k_{0}$ considered just above. I display these measurements in figure \ref{volprofile}\subref{volprofilek0}. 
Rescaling $\langle N_{2}^{\mathrm{SL}}\rangle$ by a function $f_{\mathrm{SL}}(k_{0})$ and $\tau$ by a function $f_{\mathrm{TL}}(k_{0})$, I confirm the finding of Ambj\o rn, Jurkiewicz, and Loll: as the plot in figure \ref{volprofile}\subref{scaledvolprofilek0} shows, for different values of $k_{0}$, $\langle N_{2}^{\mathrm{SL}}(\tau)\rangle$ differs only by rescaling by the functions $f_{\mathrm{SL}}(k_{0})$ and $f_{\mathrm{TL}}(k_{0})$.  

\begin{figure}[h]
\centering
\subfigure[]{
\includegraphics[width=0.45\linewidth]{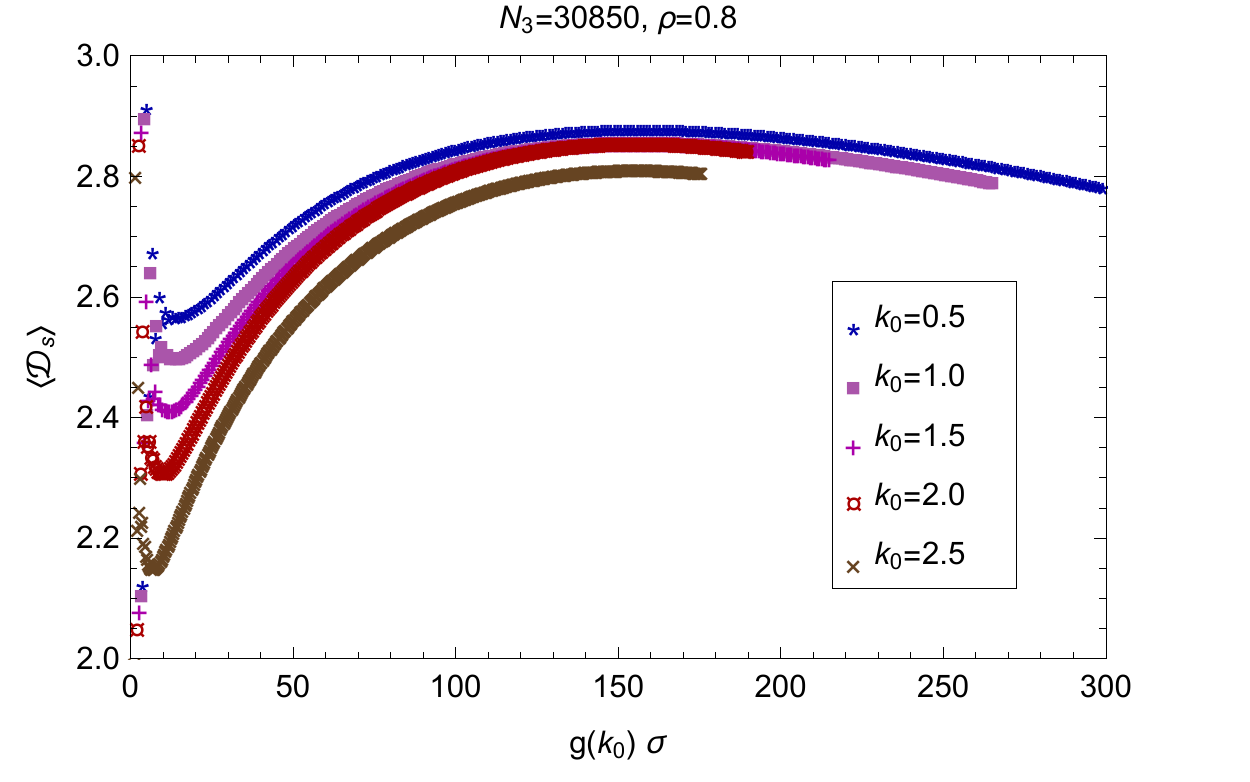}
\label{scaledspecdimvaryingk0}}
\subfigure[]{
\includegraphics[width=0.45\linewidth]{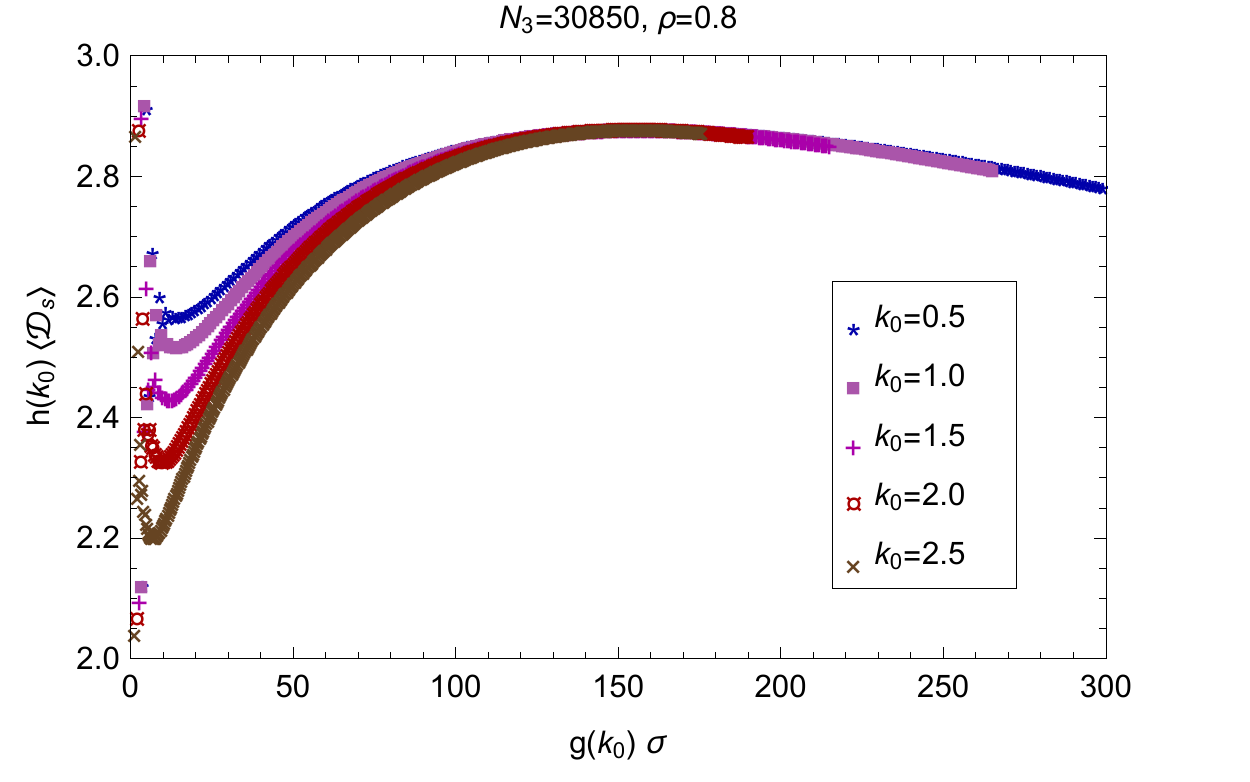}
\label{doublescaledspecdimvaryingk0}}
\caption{\subref{scaledspecdimvaryingk0} The ensemble average spectral dimension $\langle \mathcal{D}_{\mathfrak{s}}\rangle$ as a function of the diffusion time $g(k_{0})\,\sigma$, scaled by the function $g(k_{0})$, for five values of the bare coupling $k_{0}$ for $N_{3}=30850$, $T=64$, and $\rho=0.8$. \subref{doublescaledspecdimvaryingk0} The ensemble average spectral dimension $h(k_{0})\,\langle \mathcal{D}_{\mathfrak{s}}\rangle$, scaled by the function $h(k_{0})$, as a function of the diffusion time $g(k_{0})\,\sigma$, scaled by the function $g(k_{0})$, for five values of the bare coupling $k_{0}$ for $N_{3}=30850$, $T=64$, and $\rho=0.8$.}
\label{scaledspecdimk0}
\end{figure}

The numerical measurements of $\langle\mathcal{D}_{\mathfrak{s}}(\sigma)\rangle$ depicted in figures \ref{specdimvaryingk0sep} and \ref{specdimvaryingk0} serve to further test the conjecture of Ambj\o rn, Jurkiewicz, and Loll. %appear to contradict their conjecture. 
Since the spectral dimension is dimensionless, I do not expect to its value to scale with the hypothetical scale set by $k_{0}$; since the diffusion time is dimensionful, I do expect its value to scale with the hypothetical scale set by $k_{0}$. Accordingly, rescaling $\sigma$ by a function $g(k_{0})$, I plot $\langle\mathcal{D}_{\mathfrak{s}}\rangle$ as a function of $g(k_{0})\,\sigma$ in figure \ref{scaledspecdimk0}\subref{scaledspecdimvaryingk0}. 
Rescaling $\sigma$ alone fails to result in complete overlapping of $\langle\mathcal{D}_{\mathfrak{s}}(\sigma)\rangle$ for different values of $k_{0}$. Of course, I expected this failure on the basis of the plot in figure \ref{sigmamaxvaryingk0}\subref{sigmamaxk0}. Rescaling also $\langle\mathcal{D}_{\mathfrak{s}}\rangle$ by a function $h(k_{0})$, I plot $h(k_{0})\,\langle\mathcal{D}_{\mathfrak{s}}\rangle$ as a function of $g(k_{0})\,\sigma$ in figure \ref{scaledspecdimk0}\subref{doublescaledspecdimvaryingk0}. Rescaling both $\langle\mathcal{D}_{\mathfrak{s}}\rangle$ and $\sigma$ by functions of $k_{0}$ results in complete overlapping of $\langle\mathcal{D}_{\mathfrak{s}}(\sigma)\rangle$ for different values of $k_{0}$ on intermediate to large scales but not on small scales. Numerical measurements of $\langle\mathcal{D}_{\mathfrak{s}}(\sigma)\rangle$ therefore indicate that $k_{0}$ sets the overall scale of the quantum geometry on sufficiently large scales but that $k_{0}$ does not set the overall scale of the quantum geometry on sufficiently small scales. While this conclusion is consistent with the findings of Ambj\o rn, Jurkiewicz, and Loll, confirmed just above, since $\langle N_{2}^{\mathrm{SL}}(\tau)\rangle$ is a large-scale observable, this conclusion necessitates a refinement of these authors' conjecture. The breakdown of their conjecture is not unexpected: as evidenced by the dynamics of $\langle\mathcal{D}_{\mathfrak{s}}(\sigma)\rangle$, at least three scales---the rate of dynamical dimensional reduction, the scale associated with $\sigma_{\mathrm{max}}$, and the rate of large-scale decay of $\langle\mathcal{D}_{\mathfrak{s}}(\sigma)\rangle$---characterize the quantum geometry within phase C, of which only the last coincides with the overall scale of $\langle N_{2}^{\mathrm{SL}}(\tau)\rangle$.

\section{Conclusion}\label{conclusion}

I have aimed to study comprehensively the phenomenology of the small-to-intermediate scale spectral dimension within phase C of the causal dynamical triangulations of $3$-dimensional Einstein gravity. To this end I performed %and presented 
systematic numerical measurements of the ensemble average spectral dimension $\langle\mathcal{D}_{\mathfrak{s}}(\sigma)\rangle$ as a function the diffusion time $\sigma$. I presented and analyzed these measurements in section \ref{FSS}, investigating the dependence of $\langle\mathcal{D}_{\mathfrak{s}}(\sigma)\rangle$ on the three parameters that enter into a measurement of $\langle\mathcal{D}_{\mathfrak{s}}(\sigma)\rangle$: the diffusion constant $\rho$, the number $N_{3}$ of $3$-simplices, and the bare coupling $k_{0}$. In subsection \ref{diffconstdependence} I demonstrated that $\langle\mathcal{D}_{\mathfrak{s}}(\sigma)\rangle$ depends trivially on $\rho$: $\rho$ simply rescales $\sigma$ as equation \eqref{heatequation}, describing diffusion on a Riemannian manifold, implies. In subsection \ref{N3dependence} I found that the maximal value $\langle\mathcal{D}_{\mathfrak{s}}\rangle_{\mathrm{max}}$ of $\langle\mathcal{D}_{\mathfrak{s}}(\sigma)\rangle$ is finite in the infinite volume limit but that $\langle\mathcal{D}_{\mathfrak{s}}\rangle_{\mathrm{max}}$ slightly overshoots the topological dimension of $3$. In subsection \ref{k0dependence} I found that the minimal value $\langle\mathcal{D}_{\mathfrak{s}}\rangle_{\mathrm{min}}$ of $\langle\mathcal{D}_{\mathfrak{s}}(\sigma)\rangle$---the value to which $\langle\mathcal{D}_{\mathfrak{s}}(\sigma)\rangle$ dynamically reduces---decreases with $k_{0}$ towards the AC phase transition. %This constitutes another respect in which $\langle\mathcal{D}_{\mathfrak{s}}(\sigma)\rangle$ mimics the phenomenology of the spectral dimension within phase C of the causal dynamical triangulations of $4$-dimensional Einstein gravity \cite{DNC&JJ}.
This last finding leads to the conclusion that $k_{0}$ alone does not set the scales of the quantum geometry on sufficiently small scales, counter to a conjecture of Ambj\o rn, Jurkiewicz, and Loll \cite{JA&JJ&RL3}. 

Taken together, the results of section \ref{FSS} lead to several novel insights. The findings of subsections \ref{N3dependence} and \ref{k0dependence} point to the possibility of $\langle\mathcal{D}_{\mathfrak{s}}\rangle_{\mathrm{max}}$ equaling $3$ in the infinite volume limit. While $\langle\mathcal{D}_{\mathfrak{s}}\rangle_{\mathrm{max}}$ clearly overshoots $3$ in the infinite volume limit for $k_{0}=1.0$, $\langle\mathcal{D}_{\mathfrak{s}}\rangle_{\mathrm{max}}$ decreases as $k_{0}$ approaches the AC phase transition for fixed $N_{3}$. By adjusting $k_{0}$ to a value near $k_{0}^{c}$, I propose that $\langle\mathcal{D}_{\mathfrak{s}}\rangle_{\mathrm{max}}$ would be precisely tuned to a value of $3$ in the infinite volume limit. The quantum geometry of phase C would then exhibit exact (spectral) $3$-dimensionality on the intermediate scale associated with $\sigma_{\mathrm{max}}$. % up to its largest scales. %compensate for slight overshoot by adjusting coupling?
A finite-size scaling analysis of $\langle\mathcal{D}_{\mathfrak{s}}(\sigma)\rangle$ for $k_{0}\lesssim k_{0}^{c}$ would readily test this proposal.

\begin{figure}[h]
\centering
\includegraphics[width=0.45\linewidth]{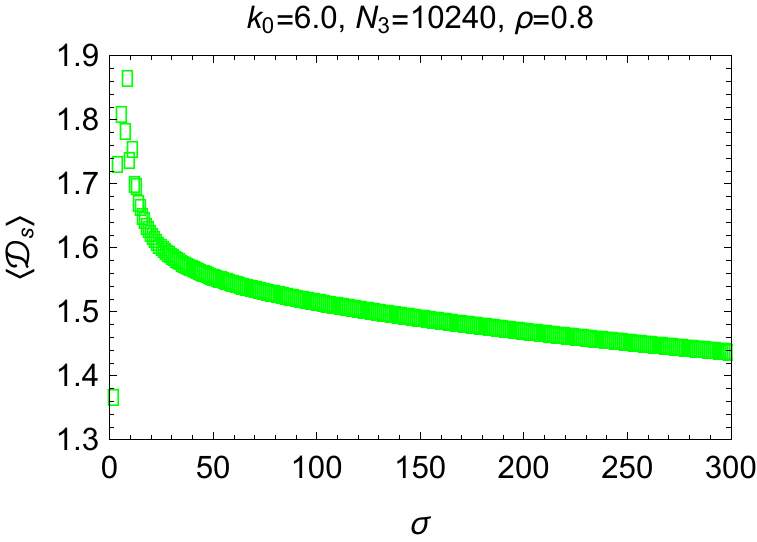}
\caption{The ensemble average spectral dimension $\langle\mathcal{D}_{\mathfrak{s}}\rangle$ as a function of the diffusion time $\sigma$ for $k_{0}=6.0$, $N_{3}=10240$, $T=64$, and $\rho=0.8$.}
\label{specdimphaseA}
\end{figure}

The findings of subsection \ref{k0dependence} %displayed in figures \ref{specdimvaryingk0}, \ref{specdimmaxvaryingk0}, \ref{sigmamaxvaryingk0}, and \ref{specdimminvaryingk0} 
point to consistency with the spectral dimension within phase A. %AC phase transition in the following manner. %the known behavior of $\langle\mathcal{D}_{\mathfrak{s}}(\sigma)\rangle$ within phase A: 
As Anderson \emph{et al} first reported \cite{CA&SC&JHC&PH&RKK&PZ}, %(for $k_{0}=6.0$, $N_{3}=10240$, and $\rho=1$), 
$\langle\mathcal{D}_{\mathfrak{s}}(\sigma)\rangle$ within phase A is an approximately constant function of $\sigma$ with a value of approximately $\frac{3}{2}$. Coumbe and Jurkiewicz subsequently measured $\langle\mathcal{D}_{\mathfrak{s}}(\sigma)\rangle$ within phase A of the causal dynamical triangulations of $4$-dimensional Einstein gravity, finding that $\langle\mathcal{D}_{\mathfrak{s}}(\sigma)\rangle$ is an approximately constant function of $\sigma$ with a value of approximately $\frac{4}{3}$ \cite{DNC&JJ}. In fact, both of these measurements show $\langle\mathcal{D}_{\mathfrak{s}}(\sigma)\rangle$ decreasing very slowly with $\sigma$, presumably owing to the compact topology of each causal triangulation. I have measured $\langle\mathcal{D}_{\mathfrak{s}}(\sigma)\rangle$ for four values of $k_{0}$ within phase A (including the value of $k_{0}$ considered by Anderson \emph{et al}). %$k_{0}=4.0$, $k_{0}=5.0$, $k_{0}=6.0$, and $k_{0}=7.0$ for $N_{3}=10240$ and $\rho=0.8$. 
As these four measurements reveal very little variation of $\langle\mathcal{D}_{\mathfrak{s}}(\sigma)\rangle$ with $k_{0}$, I plot in figure \ref{specdimphaseA} just the case of $k_{0}=6.0$, %Within phase A of the causal dynamical triangulation of $3$-dimensional Einstein gravity, the spectral dimension is a constant function of $\sigma$ (up to measurement errors) with a value of approximately $\frac{3}{2}$ \cite{CA&SC&JHC&PH&RKK&PZ}; within phase A 
which shows consistency with the measurement of \cite{CA&SC&JHC&PH&RKK&PZ}. Since Anderson \emph{et al} and I measured $\langle\mathcal{D}_{\mathfrak{s}}(\sigma)\rangle$ for $k_{0}=6.0$ at $N_{3}=10240$, the value of approximately $\frac{3}{2}$ obtained might be depressed by finite-size effects in which case $\langle\mathcal{D}_{\mathfrak{s}}(\sigma)\rangle$ might approach a value of $\frac{4}{3}$ in the infinite volume limit.

Phase A is regarded as the analogue in causal dynamical triangulations of the branched polymeric phase of Euclidean dynamical triangulations \cite{JA&JJ&RL3,JA&JJ&RL6}. Branched polymers are self-similar or scale-invariant structures as evidenced, for instance, by the constancy of their spectral dimension with the diffusion time. Ordinary branched polymers have a spectral dimension of $\frac{4}{3}$ \cite{TJ&JFW} while a novel class of branched polymers allows for a spectral dimension of $\frac{3}{2}$ \cite{JA&BD&TJ,JA&BD&TJ&GT,JA&JJ&RL6}. The spectral dimension of phase A is therefore consistent with that of branched polymers. %(4d phase diagram, analytical and numerical results in 2d) %There are particular classes of branches polymers for which $D_{\mathfrak{s}}=\frac{3}{2}$ and for which $D_{\mathfrak{s}}=\frac{4}{3}$ \cite{}. 

I now attempt to reconcile the numerical measurements of $\langle\mathcal{D}_{\mathfrak{s}}(\sigma)\rangle$ within phase A with the findings of subsection \ref{k0dependence}. Extrapolating from figure \ref{specdimmaxvaryingk0}, I hypothesize that $\langle\mathcal{D}_{\mathfrak{s}}\rangle_{\mathrm{max}}$ continues to decrease with $k_{0}$ towards a value of approximately $\frac{3}{2}$ at the AC phase transition. %This extrapolation, while well supported by figures \ref{specdimmaxvaryingk0} and \ref{specdimphaseA}, could prove erroneous: the decrease in $\langle\mathcal{D}_{\mathfrak{s}}\rangle_{\mathrm{max}}$ for $k_{0}$ near the AC phase transition could be a finite-size effect pronounced by proximity to the phase transition in which case $\langle\mathcal{D}_{\mathfrak{s}}\rangle_{\mathrm{max}}$ could jump discontinuously from a value of approximately 3 to a value of approximately $\frac{3}{2}$ across the AC phase transition. 
Extrapolating from figure \ref{sigmamaxvaryingk0}, I hypothesize that $\sigma_{\mathrm{max}}$ diverges as $k_{0}$ approaches $k_{0}^{c}$. %This extrapolation is well supported by figures \ref{sigmamaxvaryingk0} and \ref{specdimphaseA}. 
Extrapolating from figure \ref{specdimminvaryingk0}, I hypothesize that $\langle\mathcal{D}_{\mathfrak{s}}\rangle_{\mathrm{min}}$ continues to decrease with $k_{0}$ towards a value of approximately $\frac{3}{2}$ at the AC phase transition. Equivalently, I propose that $\langle\mathcal{D}_{\mathfrak{s}}(\sigma)\rangle$ changes continuously from its characteristic behavior in phase C, depicted in figures \ref{diffconstdependence10}, \ref{diffconstdependence10together}, \ref{specdimN3}, \ref{specdimvaryingk0sep}, and \ref{specdimvaryingk0}, to its characteristic behavior in phase A, depicted in figure \ref{specdimphaseA}, as $k_{0}$ approaches $k_{0}^{c}$.

This hypothesized consistency of $\langle\mathcal{D}_{\mathfrak{s}}(\sigma)\rangle$ across the AC phase transition suggests a physical explanation for the dynamical reduction of $\langle\mathcal{D}_{\mathfrak{s}}(\sigma)\rangle$ observed within phase C: on sufficiently small scales the quantum geometry of phase C is branched polymeric. In particular, the decrease in $\langle\mathcal{D}_{\mathfrak{s}}\rangle_{\mathrm{min}}$ with $k_{0}$ would halt at a value of approximately $\frac{3}{2}$, fixed by the branched polymeric nature of the quantum geometry on sufficiently small scales. This explanation's viability clearly rests on the findings of future investigations, for instance, of the type performed in \cite{JA&JJ&RL6}. %but I choose to advance this explanation on the basis of the above findings. 
This explanation itself suggests that ensembles of causal triangulations closer to the AC phase transition probe smaller physical scales for fixed $N_{3}$. An analysis of the type performed in \cite{JHC&KL&JMM} might test this suggestion. %or that the branched polymeric behavior becomes manifest on larger scales for larger values of $k_{0}$. 

This physical explanation of dynamical dimensional reduction also harmonizes with the aim of causal dynamical triangulations. % to define a quantum theory of gravity as the continuum limit. 
As I briefly discussed at the end of section \ref{CDT}, %within the causal dynamical triangulations approach 
one hopes to remove the regularization by causal triangulations %, is to define a quantum theory of gravity as the 
by taking a continuum limit at a nontrivial ultraviolet fixed point. The theory corresponding to a fixed point is scale-invariant, and, as I noted just above, branched polymers are scale-invariant structures. % as evidenced, for instance, by the constancy of their spectral dimensions with the diffusion time.
Branched polymeric quantum geometry on sufficiently small scales could thus signal the existence of the desired fixed point.
%The $\sigma$-dependence of $\langle \mathcal{D}_{s}(\sigma)\rangle$ scales anomalously with $N_{3}$ for $\sigma<\sigma_{\mathrm{max}}$. How does the $\sigma$-dependence of $\langle \mathcal{D}_{s}(\sigma)\rangle$ scale with $N_{3}$ for $\sigma>\sigma_{\mathrm{max}}$? In particular, does canonical scaling of the $\sigma$-dependence of $\langle\mathcal{D}_{s}(\sigma)\rangle$ emerge for some $\sigma>\sigma_{\mathrm{max}}$ as suggested in \cite{DB&JH}?
%How does $\langle \mathcal{D}_{s}(\sigma)\rangle$ depend on $k_{0}$ for fixed $\rho$ and $N_{3}$? 

To determine whether a continuum limit exists, one typically performs a renormalization group analysis. %As I briefly explained at the end of section \ref{CDT}, fixed points are typically located along second-order (or higher-order) phase transitions. 
Specifically, one looks for renormalization group trajectories connected to a candidate fixed point location along which the lattice spacing approaches zero (in physical units) as the trajectories approach this location. If such trajectories exist, then one can take a continuum limit by tuning the bare couplings to the fixed point. Such an analysis assumes that one knows how to delineate renormalization group flows. %, usually accomplished \emph{via} a coarse-graining scheme. 
In most contexts---specifically, those in which there is a fixed spacetime---this assumption is completely justified: there exist well-established techniques for delineating renormalization group flows, for instance, coarse-graining transformations or flow equations. In the context of nonperturbative quantum theories of gravity---in which case there is no fixed spacetime---delineating renormalization group trajectories is considerably more difficult \cite{JHC5}. How does one delineate a renormalization group trajectory in the latter context? As Ambj\o rn \emph{et al} and Cooperman clearly explain \cite{JA&AG&JJ&AK&RL,JHC}, one must return to the defining characteristic of a renormalization group trajectory: the physics described by the succession of effective theories along a renormalization group trajectory remains fixed, merely being probed over different intervals of scales. Accordingly, one delineates a renormalization group trajectory by following fixed values of sufficiently many physical observables through the space of effective theories as the interval of scales being probed changes systematically. Of course, coarse-graining transformations and flow equations employ this definition, but  their presentations and implementations do not always make this methodology apparent. How does one change the interval of scales being probed in the latter context? At least within causal dynamical triangulations, as Ambj\o rn \emph{et al} \cite{JA&AG&JJ&AK&RL} and Cooperman also clearly explain \cite{JA&AG&JJ&AK&RL,JHC}, one %can accomplish this by 
directly simulates ensembles of causal triangulations intrinsically characterized by different intervals of scales. 

Ambj\o rn \emph{et al} and Cooperman both proposed renormalization group schemes for causal dynamical triangulations on this basis; however, their schemes were too deficient in physical observables to allow for unambiguous delineation of renormalization group trajectories \cite{JA&AG&JJ&AK&RL,JHC}. To overcome this deficiency, Ambj\o rn \emph{et al} invoked an additional assumption, a constant lattice spacing \cite{JA&AG&JJ&AK&RL}, whereas Cooperman proposed employing an additional physical observable, the spectral dimension \cite{JHC}. The value of the spectral dimension at a fixed physical scale---as opposed to its value at a fixed diffusion time---is a physical observable. One could therefore use the spectral dimension in delineating renormalization group trajectories provided that one could also determine the physical scale at which the spectral dimension takes a specified value. Making this determination proves rather nontrivial, but I am currently pursuing this goal \cite{JHC&MD}. Since one simulates different ensembles of causal triangulations to change the interval of scales being probed, using the spectral dimension to delineate renormalization group trajectories translates into determining the circumstances under which measurements of $\langle \mathcal{D}_{s}(\sigma)\rangle$ on different ensembles probe the same spectral dimension. Understanding the dependence of $\langle \mathcal{D}_{s}(\sigma)\rangle$ on the number of $D$-simplices and the bare couplings is crucial to making this determination. %I am currently pursuing this analysis.

\section*{Acknowledgements} 

I thank Christian Anderson, Jonah Miller, and especially Rajesh Kommu for allowing me to employ parts of their codes. I am indebted to Jonah Miller for providing computing facilities on which to run some of my codes. I also thank Jan Ambj\o rn, Steve Carlip, Daniel Coumbe, Hal Haggard, Joe Henson, and Renate Loll for useful discussions. I acknowledge the hospitality of the Physics Program of Bard College where I completed much of this research.

\appendix

\section{Estimates and errors}\label{appendix}

I estimate the ensemble average spectral dimension $\langle\mathcal{D}_{\mathfrak{s}}(\sigma)\rangle$ and the error $\epsilon[\langle\mathcal{D}_{\mathfrak{s}}(\sigma)\rangle]$ in $\langle\mathcal{D}_{\mathfrak{s}}(\sigma)\rangle$ as follows \cite{JHC2}. Since the number $N_{3}$ of $3$-simplices is on the order of $10^{5}$ or $10^{6}$ for the causal triangulations $\mathcal{T}_{c}$ that I have simulated, I cannot feasibly compute the return probability $\mathcal{P}_{\mathcal{T}_{c}}(\sigma)$ by summing the heat kernel elements $\mathcal{K}_{\mathcal{T}_{c}}(\mathsf{s},\mathsf{s},\sigma)$ over every $3$-simplex $\mathsf{s}$ in $\mathcal{T}_{c}$. I instead estimate $\mathcal{P}_{\mathcal{T}_{c}}(\sigma)$ by sampling a subset of $M$ randomly selected $3$-simplices $\mathsf{s}_{m}$:
\begin{equation}
\mathcal{P}_{\mathcal{T}_{c}}^{(M)}(\sigma)=\frac{1}{M}\sum_{\mathsf{s}_{m}\in\mathcal{T}_{c}}\mathcal{K}_{\mathcal{T}_{c}}(\mathsf{s}_{m},\mathsf{s}_{m},\sigma).
\end{equation}
I then estimate the spectral dimension $\mathcal{D}_{\mathfrak{s}}^{(\mathcal{T}_{c})}(\sigma)$ as follows:
\begin{equation}
\mathcal{D}_{\mathfrak{s}}^{(\mathcal{T}_{c},M)}(\sigma)=-\frac{\sigma}{\mathcal{P}_{\mathcal{T}_{c}}^{(M)}(\sigma)}\left[\mathcal{P}_{\mathcal{T}_{c}}^{(M)}(\sigma+1)-\mathcal{P}_{\mathcal{T}_{c}}^{(M)}(\sigma-1)\right].
\end{equation}
By sampling a subset of $M$ $3$-simplices, I induce an error $\epsilon[\mathcal{D}_{\mathfrak{s}}^{(\mathcal{T}_{c},M)}(\sigma)]$ in $\mathcal{D}_{\mathfrak{s}}^{(\mathcal{T}_{c},M)}(\sigma)$. I estimate $\epsilon[\mathcal{D}_{\mathfrak{s}}^{(\mathcal{T}_{c},M)}(\sigma)]$ as follows. I compute the $M$ jackknife estimates of $\mathcal{P}_{\mathcal{T}_{c}}^{(M)}(\sigma)$,
\begin{equation}
\mathcal{P}_{\mathcal{T}_{c}}^{(M_{l})}(\sigma)=\frac{1}{M-1}\sum_{\substack{\mathsf{s}_{m}\in\mathcal{T}_{c} \\ m\neq l}}\mathcal{K}(\mathsf{s}_{m},\mathsf{s}_{m},\sigma),
\end{equation}
and the $M$ corresponding jackknife estimates of $\mathcal{D}_{\mathfrak{s}}^{(\mathcal{T}_{c},M)}(\sigma)$,
\begin{equation}
\mathcal{D}_{\mathfrak{s}}^{(\mathcal{T}_{c},M_{l})}(\sigma)=-\frac{\sigma}{\mathcal{P}_{\mathcal{T}_{c}}^{(M_{l})}(\sigma)}\left[\mathcal{P}_{\mathcal{T}_{c}}^{(M_{l})}(\sigma+1)-\mathcal{P}_{\mathcal{T}_{c}}^{(M_{l})}(\sigma-1)\right],
\end{equation}
both for $l\in\{1,\ldots,M\}$. I take $\epsilon[\mathcal{D}_{\mathfrak{s}}^{(\mathcal{T}_{c},M)}(\sigma)]$ as the jackknife error in $\mathcal{D}_{\mathfrak{s}}^{(\mathcal{T}_{c},M)}(\sigma)$:
\begin{equation}
\epsilon[\mathcal{D}_{\mathfrak{s}}^{(\mathcal{T}_{c},M)}(\sigma)]=\sqrt{\frac{M-1}{M}\sum_{l=1}^{M}\left[\mathcal{D}_{\mathfrak{s}}^{(\mathcal{T}_{c},M_{l})}(\sigma)-\mathcal{D}_{\mathfrak{s}}^{(\mathcal{T}_{c},M)}(\sigma)\right]^{2}}.
\end{equation}
Since the number of causal triangulations at fixed $N_{3}$ grows tremendously with $N_{3}$, I can only feasibly simulate a finite number $N(\mathcal{T}_{c})$ of causal triangulations. I instead estimate the ensemble average spectral dimension $\langle\mathcal{D}_{\mathfrak{s}}(\sigma)\rangle$ as follows:
\begin{equation}
\langle\mathcal{D}_{\mathfrak{s}}^{(M)}(\sigma)\rangle=\frac{1}{N(\mathcal{T}_{c})}\sum_{j=1}^{N(\mathcal{T}_{c})}\mathcal{D}_{\mathfrak{s}}^{(\mathcal{T}_{c}^{(j)},M)}(\sigma).
\end{equation}
By sampling a subset of $N(\mathcal{T}_{c})$ causal triangulations, I induce an error $\epsilon[\langle\mathcal{D}_{\mathfrak{s}}(\sigma)\rangle]$ in $\langle\mathcal{D}_{\mathfrak{s}}(\sigma)\rangle$. I take $\epsilon[\langle\mathcal{D}_{\mathfrak{s}}(\sigma)\rangle]$ as the standard error in $\langle\mathcal{D}_{\mathfrak{s}}(\sigma)\rangle$: %There are two sources of error in $\langle D_{s}(\sigma)\rangle$: error induced by sampling a subset of $K$ $3$-simplices for each causal triangulations, and error induced by sampling a subset of causal triangulations in the ensemble. For each causal triangulations in an ensemble, I first compute the $K$ jackknife estimates of $\mathcal{P}_{\mathcal{T}_{c}}^{(K)}(\sigma)$,
\begin{equation}
\epsilon[\langle\mathcal{D}_{\mathfrak{s}}(\sigma)\rangle]=\sqrt{\frac{1}{N(\mathcal{T}_{c})[N(\mathcal{T}_{c})-1]}\sum_{j=1}^{N(\mathcal{T}_{c})}\left[\mathcal{D}_{\mathfrak{s}}^{(\mathcal{T}_{c}^{(j)},M)}(\sigma)-\langle\mathcal{D}_{\mathfrak{s}}^{(M)}(\sigma)\rangle\right]^{2}}
\end{equation}
I take the total error as
\begin{equation}
\langle\epsilon[D_{s}^{(M)}(\sigma)]\rangle+\epsilon[\langle D_{s}(\sigma)\rangle]
\end{equation}
in which
\begin{equation}
\langle\epsilon[\mathcal{D}_{\mathfrak{s}}^{(M)}(\sigma)]\rangle=\frac{1}{N(\mathcal{T}_{c})}\sqrt{\sum_{j=1}^{N(\mathcal{T}_{c})}\epsilon^{2}[\mathcal{D}_{\mathfrak{s}}^{(\mathcal{T}_{c}^{(j)},M)}(\sigma)]},
\end{equation}
the error propagated to $\langle\mathcal{D}_{\mathfrak{s}}(\sigma)\rangle$ from $\mathcal{D}_{\mathfrak{s}}^{(\mathcal{T}_{c},M)}(\sigma)$ in computing the ensemble average.


\begin{thebibliography}{99}

\bibitem{JA&DNC&JGS&AG&JJ} J. Ambj\o rn, D. N. Coumbe, J. Gizbert-Studnicki, A. G\"{o}rlich, and J. Jurkiewicz. ``New higher-order transition in causal dynamical triangulations." \emph{Physical Review D} 95 (2017) 124029.

\bibitem{JA&DNC&JGS&AG&JJ&NK&RL} J. Ambj\o rn, D. N. Coumbe, J. Gizbert-Studnicki, A. G\"{o}rlich, J. Jurkiewicz, N. Klitgaard, and R. Loll. ``Characteristics of the new phase in CDT." \emph{European Physical Journal C} 77 (2017) 152.

\bibitem{JA&DNC&JGS&JJ} J. Ambj\o rn, D. N. Coumbe, J. Gizbert-Studnicki, and J. Jurkiewicz. ``Searching for a continuum limit of causal dynamical triangulation quantum gravity." \emph{Physical Review D} 93 (2016) 104032. %arXiv: hep-th/1603.02076.

\bibitem{JA&BD&TJ} J. Ambj\o rn, B. Durhuus, B. Durhuus, and T. Jonsson. ``A solvable 2D quantum gravity model with $\gamma>0$." \emph{Modern Physics Letters A} 9 (1994) 1221. 

\bibitem{JA&BD&TJ&GT} J. Ambj\o rn, B. Durhuus, T. Jonsson, and G. Thorleifsson. ``Matter fields with $c>1$ coupled to 2d gravity." \emph{Nuclear Physics B} 398 (1993) 568.

\bibitem{JA&AG&JJ&AK&RL} J. Ambj{\o}rn, A. G\"{o}rlich, J. Jurkiewicz, A. Kreienbuehl, and R. Loll. ``Renormalization group flow in CDT." \emph{Classical and Quantum Gravity} 31 (2014) 165003. 

\bibitem{JA&SJ&JJ&RL1} J. Ambj{\o}rn, S. Jordan, J. Jurkiewicz, and R. Loll. ``Second-Order Phase Transition in Causal Dynamical Triangulations." \emph{Physical Review Letters} 107 (2011) 211303.

\bibitem{JA&SJ&JJ&RL2} J. Ambj{\o}rn, S. Jordan, J. Jurkiewicz, and R. Loll. ``Second- and first-order phase transitions in causal dynamical triangulations." \emph{Physical Review D} 85 (2012) 124044.

\bibitem{JA&JJ&RL1} J. Ambj{\o}rn, J. Jurkiewicz, and R. Loll. ``Non-perturbative Lorentzian Path Integral for Gravity." \emph{Physical Review Letters} 85 (2000) 347.

\bibitem{JA&JJ&RL2} J. Ambj{\o}rn, J. Jurkiewicz, and R. Loll. ``Dynamically triangulating Lorentzian quantum gravity." \emph{Nuclear Physics B} 610 (2001) 347.

\bibitem{JA&JJ&RL3} J. Ambj{\o}rn, J. Jurkiewicz, and R. Loll. ``Nonperturbative 3d Lorentzian Quantum Gravity." \emph{Physical Review D} 64 (2001) 044011.

\bibitem{JA&JJ&RL7} J. Ambj{\o}rn, J. Jurkiewicz, and R. Loll. ``The Spectral Dimension of the Universe is Scale Dependent."  \emph{Physical Review Letters} 95 (2005) 171301.

\bibitem{JA&JJ&RL6} J. Ambj{\o}rn, J. Jurkiewicz, and R. Loll. ``Reconstructing the universe." \emph{Physical Review D} 72 (2005) 064014.

\bibitem{JA&RL} J. Ambj\o rn and R. Loll. ``Non-perturbative Lorentzian quantum gravity, causality, and topology change." \emph{Nuclear Physics B} 536 (1998) 407.

\bibitem{CA&SC&JHC&PH&RKK&PZ} C. Anderson, S. Carlip, J. H. Cooperman, P. Ho\v{r}ava, R. K. Kommu, and P. Zulkowski. ``Quantizing Ho\v{r}ava-Lifshitz gravity via causal dynamical triangulations." \emph{Physical Review D} 85 (2012) 049904.

\bibitem{DB&JH} D. Benedetti and J. Henson.  ``Spectral geometry as a probe of quantum spacetime." \emph{Physical Review D} 80 (2009) 124036.

\bibitem{DB&JH2} D. Benedetti and J. Henson. ``Spacetime condensation in $(2+1)$-dimensional CDT from a Ho\v{r}ava-Lifshitz minisuperspace model." \emph{Classical and Quantum Gravity} 32 (2015) 215007. 

\bibitem{SC} S. Carlip. ``Dimension and dimensional reduction in quantum gravity." \emph{Classical and Quantum Gravity} 34 (2017) 193001.

\bibitem{JHC5} J. H. Cooperman. ``Renormalization of lattice-regularized quantum gravity models I. General considerations." arXiv: 1410.0026.

\bibitem{JHC2} J. H. Cooperman. ``Scale-dependent homogeneity measures for causal dynamical triangulations." \emph{Physical Review D} 90 (2014) 124053.

\bibitem{JHC} J. H. Cooperman. ``On a renormalization group scheme for causal dynamical triangulations." \emph{General Relativity and Gravitation} 48 (2016) 1. 

\bibitem{JHC4} J. H. Cooperman. ``Comments on `Searching for a continuum limit in CDT quantum gravity'." arXiv: 1604.01798

%\bibitem{JHC3} J. H. Cooperman. ``Tracing renormalization group flows with the spectral dimension in causal dynamical triangulations." In preparation. 

\bibitem{JHC&MD} J. H. Cooperman and M. Dorghabekov. ``Setting the scale of dynamical dimensional reduction in causal dynamical triangulations." In preparation. 

\bibitem{JHC&KL&JMM} J. H. Cooperman, K. Lee, and J. M. Miller. ``A second look at transition amplitudes in $(2+1)$-dimensional causal dynamical triangulations." \emph{Classical and Quantum Gravity} 34 (2017) 115008.

\bibitem{JHC&JMM} J. H. Cooperman and J. M. Miller. ``A first look at transition amplitudes in $(2+1)$-dimensional causal dynamical triangulations." \emph{Classical and Quantum Gravity} 31 (2014) 035012.

\bibitem{DNC&JJ} D. N. Coumbe and J. Jurkiewicz. ``Evidence for asymptotic safety from dimensional reduction in causal dynamical triangulations." \emph{Journal of High Energy Physics} 03 (2015) 151.

\bibitem{TJ&JFW} T. Jonsson and J. F. Wheater. ``The spectral dimension of the branched polymers phase of two-dimensional quantum gravity." \emph{Nuclear Physics B} 515 (1998) 549.

\bibitem{RK} R. K. Kommu. ``A validation of causal dynamical triangulations." \emph{Classical and Quantum Gravity} 29 (2012) 105003.

\end{thebibliography}
\end{document}